\documentclass[showpacs,aps,prd,eqsecnum,floats]{revtex4}

\newif\ifpdf\ifx\pdfoutput\undefined\pdffalse\else\pdfoutput=1\pdftrue\fi
\usepackage{graphicx}
\usepackage{amsfonts}
\usepackage{subfigure}

\graphicspath{{./graphics/}}
\ifpdf\usepackage{hyperref}\else\fi
\newcommand{\be}{\begin{equation}}
\newcommand{\ee}{\end{equation}}

\newcommand{\B}{\beta}
\newcommand{\delT}{\Delta T}
\newcommand{\delTobs}{\widetilde{\Delta T}}

\newcommand{\lp}{{l^\prime}}
\newcommand{\mpp}{m^{\prime\prime}}
\newcommand{\p}{\prime}
\newcommand{\q}{\mathbf{\hat{q}}}
\newcommand{\qp}{\mathbf{\hat{q}^\prime}}

\newcommand{\cclobs}{\langle \widetilde{\mathcal{C}}_l \rangle}
\newcommand{\cclobsp}{\langle \widetilde{\mathcal{C}}_l^\prime \rangle}
\newcommand{\ccov}{\mbox{Cov}(\widetilde{\mathcal{C}}_l,\widetilde{\mathcal{C}}_\lp)}
\newcommand{\ccovp}{\mbox{Cov}(\widetilde{\mathcal{C}}_l^\p,\widetilde{\mathcal{C}}_\lp^\p)}
\newcommand{\club}{\widetilde{\mathcal{C}}_l^{UB}}
\newcommand{\clubp}{\widetilde{\mathcal{C}}_{l^\prime}^{UB}}

\def\lsim{\mathrel{\rlap{\lower4pt\hbox{\hskip1pt$\sim$}}
    \raise1pt\hbox{$<$}}}                % less than or approx. symbol
\def\gsim{\mathrel{\rlap{\lower4pt\hbox{\hskip1pt$\sim$}}
    \raise1pt\hbox{$>$}}}                % greater than or approx. symbol

\begin{document}

\preprint{IUCAA xx/2004}

\title{CMB power spectrum estimation using noncircular beams}
\author{Sanjit Mitra \email{sanjit@iucaa.ernet.in}}
\author{Anand S. Sengupta \email{anandss@iucaa.ernet.in}}
\author{Tarun Souradeep \email{tarun@iucaa.ernet.in}}

\affiliation{Inter-University Centre for Astronomy and Astrophysics,\\
Post Bag 4, Ganeshkhind, Pune 411007, India.}

\begin{abstract}

The measurements of the angular power spectrum of the Cosmic Microwave
Background (CMB) anisotropy has proved crucial to the emergence of
cosmology as a precision science in recent years.  In this remarkable
data rich period, the limitations to precision now arise from the the
inability to account for finer systematic effects in data analysis.
The non-circularity of the experimental beam has become progressively
important as CMB experiments strive to attain higher angular
resolution and sensitivity.  We present an analytic framework for
studying the leading order effects of a non-circular beam on the CMB
power spectrum estimation. We consider a non-circular beam of fixed
shape but variable orientation. We compute the bias in the pseudo-$C_l$
power spectrum estimator and then construct an unbiased estimator
using the bias matrix.  The covariance matrix of the unbiased
estimator is computed for smooth, non-circular beams.  Quantitative
results are shown for CMB maps made by a \emph{hypothetical}
experiment with a non-circular beam comparable to our fits to the WMAP
beam maps described in the appendix and uses a \emph{toy} scan
strategy. We find that significant effects on CMB power spectrum can
arise due to non-circular beam on multipoles comparable to, and
beyond, the inverse average beam-width where the pseudo-$C_l$ approach
may be the method of choice due to computational limitations of
analyzing the large datasets from current and near future CMB
experiments.

\end{abstract}

\maketitle
%%%%%%%%%%%%%%%%%%%%%%%%%%%%%%%%%%%%%%%%%%%%%%%%%%%%%%%%%%%%%%%%%%%%%%%%%%%
\section{Introduction}
\label{secintro}

A golden decade of measurements of the cosmic microwave background
anisotropy has ushered in an era of precision cosmology.  The theory
of primary CMB anisotropy is well developed and the past decade has
seen a veritable flood of data~\cite{hu_dod02,bon04}.  Increasingly
sensitive, high resolution, `full' sky measurements from space
missions, such as, the ongoing Wilkinson Microwave Anisotropy Probe
(WMAP) and, the upcoming Planck surveyor pose a stiff challenge for
current analysis techniques to realize the full potential of precise
determination of cosmological parameters.  As experiments improve in
sensitivity, the inadequacy in modeling the observational reality
start to limit the returns from these experiments.

A Gaussian model of CMB anisotropy $\Delta T(\q)$ is completely
specified by its angular two-point correlation function. In standard
cosmology, CMB anisotropy is expected to be statistically isotropic.
In spherical harmonic space, where $\Delta T(\q)= \sum_{lm}a_{lm}
Y_{lm}(\q)$, this translates to a diagonal $\langle a_{lm}
a^*_{l^\prime m^\prime}\rangle=C_{l}
\delta_{ll^\prime}\delta_{mm^\prime}$ where $C_l$, the widely used
angular power spectrum of CMB anisotropy, is a complete description of
a Gaussian CMB anisotropy.  Observationally, the angular power
spectrum being a simple, robust point statistics is the obvious first
target for cosmological observations. Theoretically, the $C_l$ are
deemed all important since the simplest inflation models predict a
Gaussian CMB anisotropy. In this case, the power spectrum provides an
economical description of the CMB anisotropy allowing easy comparison
to observations.

Accurate estimation of $C_{l}$ is arguably the foremost concern of
most CMB experiments. The extensive literature on this topic has been
summarized in a recent article~\cite{efs04}. For Gaussian,
statistically isotropic CMB sky, the $C_l$ that correspond to
covariance that maximize the multivariate Gaussian PDF of the
temperature map, $\Delta T(\q)$ is the Maximum Likelihood (ML)
solution. Different ML estimators have been proposed and implemented
on CMB data of small and modest
size~\cite{gor94,gor_hin94,gor96,gor97,max97,bjk98}. While it is
desirable to use optimal estimators of $C_l$ that obtain (or iterate
toward) the ML solution for the given data, these methods usually are
limited by the computational expense of matrix inversion that scales
as $N_d^3$ with data size $N_d$~\cite{bor99,bon99}.  Various
strategies for speeding up ML estimation have been proposed, such as,
exploiting the symmetries of the scan strategy~\cite{oh_sper99}, using
hierarchical decomposition~\cite{dor_knox01}, iterative multi-grid
method~\cite{pen03}, etc. Variants employing linear combinations of
$\Delta T(\q)$ such as $a_{lm}$ on set of rings in the sky can
alleviate the computational demands in special
cases~\cite{harm02,wan_han03}. Other promising exact power estimation
methods have been recently proposed~\cite{wan_lar03,wan04,jew02}.

However there also exist computationally rapid, sub-optimal estimators of
$C_l$. Exploiting the fast spherical harmonic transform
($\sim N_d^{3/2}$), it is possible to estimate the angular power
spectrum $C_l= \langle|a_{lm}|^2\rangle/(2l+1)$
rapidly~\cite{yu_peeb69,peeb73}. This is commonly referred to as the
pseudo-$C_l$ method~\cite{wan_hiv03}.  (Analogous approach employing fast
estimation of the correlation function $C(\q\cdot\qp)$ have also been
explored~\cite{szap01,szap_prun01}.)  It has been recently argued that
the need for optimal estimators may have been over-emphasized since
they are computationally prohibitive at large $l$ . Sub-optimal
estimators are computationally tractable and tend to be nearly optimal in
the relevant high $l$ regime. Moreover, already the data size of the
current sensitive, high resolution, `full sky' CMB experiments such as
WMAP have compelled the use of sub-optimal pseudo-$C_l$ related
methods~\cite{Bennett:2003, hin_wmap03}.  On the other hand, optimal
ML estimators can readily incorporate and account for various
systematic effects, such as non-uniform sky coverage, noise
correlations and beam asymmetries. 

In the years after the COBE-DMR observations~\cite{cobedmr92},
more sensitive measurements at higher resolution but with
limited sky coverage were made by a number of experiments~\footnote{
For a compendium of links to experiments refer to, e.g.
http://www.mpa-garching.mpg.de/$\sim$banday/CMB.html}. The effect of
incomplete (more generally, non uniform) sky coverage on the sampling
statistics of $C_l$ was the dominant concern of these experiments
such as the ground based experiment
TOCO~\cite{toco}, DASI~\cite{dasi}, CBI~\cite{cbi},
ACBAR~\cite{acbar}, and balloon based experiments
BOOMERang~\cite{boom}, MAXIMA~\cite{maxima,lee01} and
Archeops~\cite{ben03}. Comprehensive analyzes have been carried out to
tackle this problem. For example, the basic semi-analytic framework
developed~\cite{wan_hiv03} was subsequently implemented as fast,
efficient scheme for the analysis of the BOOMERang
experiment~\cite{master}.  While the non-uniform sky coverage has been
addressed in the pseudo-$C_l$ method, the other effects remain to be
incorporated.

In this paper, we initiate a similar line of research to address a
more contemporary issue that has gained relative importance in the
post WMAP~\cite{Bennett:2003} (and pre-Planck) era of CMB anisotropy
measurement with `full' sky coverage.  It has been usual in CMB data
analysis to assume the experimental beam response to be circularly
symmetric around the pointing direction.  However, any real beam
response function has deviations from circular symmetry. Even the main
lobe of the beam response of experiments are generically non-circular
(non-axisymmetric) since detectors have to be placed off-axis on the
focal plane. (Side lobes and stray light contamination add to the
breakdown of this assumption). For high sensitive experiments, the
systematic errors arising from the beam non-circularity become
progressively more important. Recent CMB experiments such as ARCHEOPS,
MAXIMA, WMAP have significantly non-circular beams.  Future
experiments like the Planck Surveyor are expected to be even more
seriously affected by non-circular beams.

Dropping the circular beam assumption leads to major complications at
every stage of the data analysis pipeline. The extent to which the
non-circularity affects the step of going from the time-stream data to
sky map is very sensitive to the scan-strategy. The beam now has an
orientation with respect to the scan path that can potentially vary
along the path. This implies that the beam function is inherently time
dependent and difficult to deconvolve. Even after a sky map is made,
the non-circularity of the effective beam affects the estimation of
the angular power spectrum, $C_l$, by coupling the $l$ modes,
typically, on scales beyond the inverse angular beam-width.

Barring few exceptions (eg., \cite{wan_gor01}), the non-circularity of
beam patterns in CMB experiments has been addressed in limited
context.  When it has not been totally ignored, one has measured with
numerical simulations the biasing effect on the power spectrum of CMB
anisotropies of neglecting the non-circularity of the beams in the
data analysis chain (see e.g., MAXIMA~\cite{lee01,whu01},
Archeops~\cite{ben03,asym_04}).  This approach only deals with the
diagonal part of the matrix relating the observed power spectrum to
the underlying power spectrum, so does not fully describe the effect
of the beam complexity on the CMB statistics.  An integrated approach
to account for the systematic effect of a non-circular beam has not
yet been developed.

In this initial work we skip over the issues related to map making and
focus on the CMB power spectrum estimation from a CMB sky map made
with an effective beam that is non-circular.  Mild deviations from
circularity can be addressed by a perturbation approach
~\cite{TR:2001,fos02}. Besides providing an elegant analytic
formalism, the approach has lead to rapid methods for computing the
window functions for CMB experiments~\cite{pyV}. In this work the
effect of beam non-circularity on the estimation of CMB power spectrum
is studied analytically using this perturbation approach.

We present a brief primer on the connection between CMB power spectrum
and the experimental window functions in section~\ref{primer}. The
section is designed to keep the paper self-contained and also serves
to set the notation for the rest of the paper. In
section~\ref{ncirwin}, we briefly review the perturbation approach for
computing the the window functions for CMB experiments with
non-circular beam~\cite{TR:2001} and also define the elliptical
Gaussian beam and its spherical transform. The bias matrix accounting
for the non-circularity of the beam for the pseudo-$C_l$ estimator of
CMB anisotropy is derived and discussed in section~\ref{secBias}. The
error-covariance for the unbiased estimator is derived in
section~\ref{secBias}. We conclude with a discussion of the results in
section~\ref{secCov}. An interesting exercise of fitting the WMAP beam
maps with an elliptical Gaussian beam profile is presented in an
appendix~\ref{wmapbeamfit}. Details of the steps leading to our
analytical results are given in Appendix~\ref{Appint}.

\section{ Window functions of  CMB experiments: a brief primer}
\label{primer}

Conventionally, the CMB temperature, $\delT(\q)$, is expressed as a
function of angular position, $\q\equiv(\theta, \phi)$, on the sky via
the spherical harmonic decomposition,
\begin{equation}
   \delT(\q) = \sum^\infty_{l = 0}
   \sum^l_{m = -l} a_{l m} Y_{l m} (\q)\,.
   \label{dT_ylm}
\end{equation}

In an idealized noise free, CMB anisotropy sky map $\delT(\q)$ made
with infinitely high resolution, the angular power spectrum is given
by
\begin{equation}
C_l \ \equiv \ \frac{1}{2l+1}
\sum_{m=-l}^l \langle |a_{lm}|^2 \rangle,
\label{defcl}
\end{equation}
where
\begin{equation} 
a_{lm} \ \equiv \ \int d\Omega_\q \, Y_{lm}^*(\q) \, \delT(\q)
\end{equation}
are the spherical harmonic transforms of the temperature deviation
field $\delT(\q)$. We introduce the scaled power spectrum
$\mathcal{C}_l \equiv (l(l+1)/2\pi)C_l$, that measures the power per
logarithmic interval of angular scale, $l$. Eliminating $a_{lm}$, we
may write,
\begin{equation} 
\mathcal{C}_l \ = \ \frac{l(l+1)}{8\pi^2} \int d\Omega_{\q_1} \int
d\Omega_{\q_2} \langle \delT(\q_1) \delT(\q_2) \rangle
P_l(\q_1\cdot\q_2), \label{eq:cl}
\end{equation}
where we have made use of the expansion of Legendre Polynomials
\begin{equation} 
P_l(\q_1\cdot\q_2) \ = \ \frac{4\pi}{2l +1} \sum_{m=-l}^l Y_{lm}^*(\q_1) Y_{lm}(\q_2). \label{eq:pl}
\end{equation}
If we assume the isotropy of the CMB sky, $\langle \delT(\q_1)
\delT(\q_2) \rangle$ should depend only on $ \q_1 \cdot
\q_2$. Therefore, we can use Legendre expansion to show that,
\begin{equation} 
\langle \delT(\q_1) \delT(\q_2) \rangle \ = \ \sum_{l=0}^\infty
\frac{2l+1}{2l(l+1)} \, \mathcal{C}_l \,
P_l(\q_1\cdot\q_2). \label{eq:corr}
\end{equation}

All CMB anisotropy experiments measure differences in CMB temperature
at different locations on the sky. A step of map-making is required to
derive the above temperature anisotropy map at each point on the
sky. Since this is a linear operation, the correlation function of the
measured quantity for a given scanning or modulation strategy can
always be expressed as linear sum of `elementary' correlations of the
temperature given in eq.~(\ref{eq:corr}).

Typically, a CMB anisotropy experiment probes a range of angular
scales characterized by a \textit{window} function
$W_l(\q,\q^\prime)$.  The window depends both on the scanning strategy
as well as the angular resolution and response of the
experiment. However, it is neater to logically separate these two
effects by expressing the window $W_l(\q,\q^\prime)$ as a sum of
`elementary' window function of the CMB anisotropy at each point of
the map~\cite{TR:2001}. In this work, we only deal with these
elementary window functions. For a given scanning/modulation strategy,
our results can be readily generalized using the representation of the
window function as sum over elementary window functions (see, {\it
e.g.,} \cite{TR:2001,pyV}). Although the quantitative results we
present in this paper refer to a scan strategy where each pixel is
visited by the beam only once, this is not a limitation of our
approach. If pixels are multiply visited by the beam with different
orientations, the correlation function still can be expressed as a sum
over appropriate elementary window functions for which all the results
we describe in this paper hold.

\subsection{Window function for circular beams}

Due to finite resolution of the instruments, the `measured'
temperature difference $\delTobs(\q)$ along the direction
$\q$ in response to the CMB anisotropy signal $\delT(\q^\p)$ is given by
\begin{equation}
\delTobs(\q) \ = \ \int d\Omega_{\q^\p} \, B(\q,\q^\p) \, \delT(\q^\p)
\label{eq:beam}
\end{equation} 
where the experimental ``Beam" response function $B(\q,\q^\p)$
describes the sensitivity of the measuring instrument at different
angles around the pointing direction. There is an additional contribution
from instrumental noise denoted by $n(\q)$ which we shall introduce
later into our final results.

The two point correlation function for a statistical
isotropic CMB anisotropy signal is 
\begin{equation}
C(\q,\q^\prime) =  \langle \delTobs(\q) \delTobs(\qp) \rangle =
   \sum^\infty_{l = 0} {(2 l + 1) \over 4\pi}\, C_l\,\,
   W_l(\q,\,\q^\prime )\,,
   \label{corre}
\end{equation}
where $C_l$ is the angular spectrum of CMB anisotropy signal and the
window function
\begin{equation} 
W_l(\q_1,\q_2) \ \equiv \ \int d\Omega_\q \int d\Omega_\qp \, B(\q_1, \q) B(\q_2, \qp) P_l(\q \cdot \qp), \label{eq:defW}
\label{windef}
\end{equation}
encodes the effect of finite resolution through the beam function.

For some experiments, the beam function may be assumed to be
circularly symmetric about the pointing direction, i.e., $B(\q,
\q^\prime) \equiv B(\q\cdot\q^\prime)$ without significantly affecting
the results of the analysis.  In any case, this assumption allows a
great simplification since the beam function can then be represented
by an expansion in Legendre polynomials as
\begin{equation}
B(\q\cdot\qp) \ = \ \frac{1}{4\pi}\,\sum_{l=0}^\infty\, (2l+1)\, B_l\,
P_l(\q\cdot\qp).
\label{eq:BthetafromBl}
\end{equation}
Consequently, it is straightforward to derive the well known simple
expression
\begin{equation}
   W_l(\q,\,\qp) \ = \ B^2_l \, P_l(\q\cdot\qp)\,,
   \label{isowine}
\end{equation}
for a circularly symmetric beam function.

\subsection{Window function for non-circular beams}
\label{ncirwin}

\begin{figure}[h]
\centering
\includegraphics[width=0.50\textwidth]{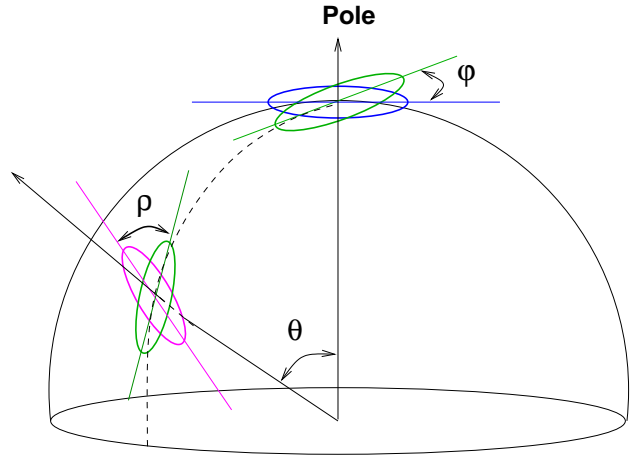}
\caption{The figure illustrates that a beam pointed in an arbitrary
direction $\q=(\theta,\phi)$, with an orientation given by the angle
$\rho(\q)$ can always be rotated to point along $\mathbf{\hat z}$ oriented with
$\rho(\mathbf{\hat z})=0$. The Euler angles of this rotation are clearly seen
to be $(\theta,\phi,\rho)$. Consequently, the beam transforms are
related through Wigner rotation matrices corresponding to the same
rotation.}
\label{fig:sphere}
\end{figure}

While some experiments may have circularly symmetric beam
functions, most experimental beams are non-circular to some
extent. The effect of non-circularity of the beam has become
progressively more relevant for experiments with higher sensitivity
and angular resolution. The most general beam response function can be
represented as
\begin{equation}
B(\mathbf{\hat z},\q) \ = \ \sum_{l=0}^\infty \sum_{m=-l}^l b_{lm}
(\mathbf{\hat z}) \, Y_{lm}(\q) \label{eq:Bexp}
\end{equation} 
by a spherical harmonic expansion when pointing along $\mathbf{\hat z}$ axis
(``North pole'' in some given astronomical coordinate system).  In case
of circularly symmetric beams, the real coefficients $B_{l} =
\sqrt{4\pi/(2l+1)}b_{l0}$.

For mild deviations, the non-circularity of the beams can be
parameterized by a set of \emph{small} quantities $\B_{lm} \equiv
b_{lm}/b_{l0}$ -- the {\em Beam Distortion Parameters} (BDP). The
smoothness of the beam response implies that at any multipole $l$, the
coefficients $\B_{lm}$ decrease sufficiently rapidly with increasing
$|m|$. In addition, for the rest of paper we assume that the beam
function has reflection symmetry about two orthogonal axes on the
(locally flat) beam plane, which ensures that the coefficients
$b_{lm}(\mathbf{\hat z})$ are real and zero for {\em odd} values of $m$. An
example of a non-circular beam with such symmetries is the elliptical
Gaussian beam.  A brief mathematical description of such beams can be
found later in this section.  In order to verify our analytical
results, we have used the elliptical Gaussian beam as a model of
non-circular beam. However, our analytic results would apply to a
general form of non-circular beam (as long as $\B_{l1}$ is zero or
sub-dominant to $\B_{l2}$).

\begin{figure}[h]
\centering
\includegraphics[width=0.53\textwidth]{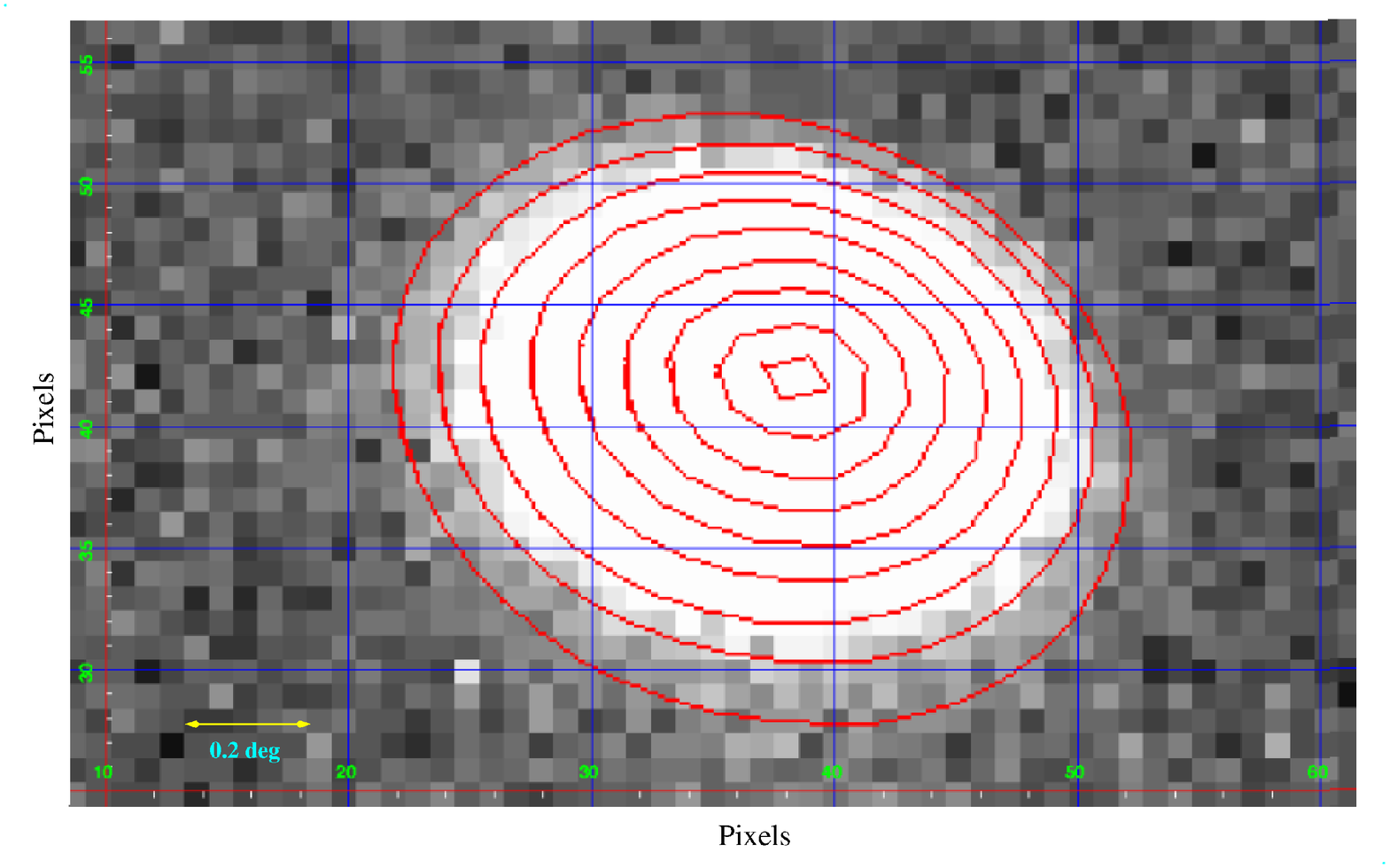}
\includegraphics[width=0.43\textwidth]{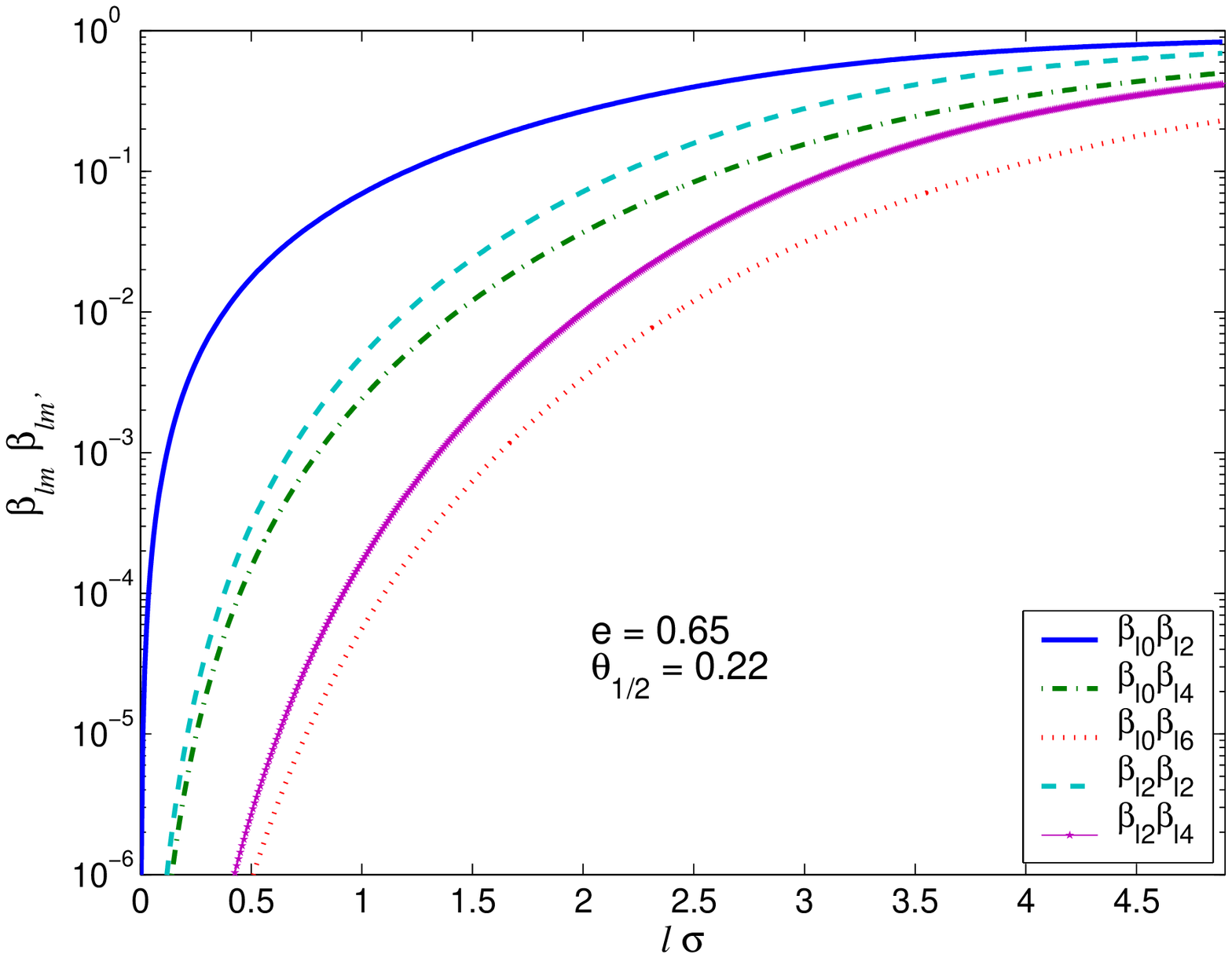}
\caption{The left panel shows the WMAP Q1 (A side) beam map overlaid with IRAF
fitted ellipses over iso-intensity contours. More details are in
Appendix~\ref{wmapbeamfit}. On the right panel, we plot the product of beam distortion
parameters for the elliptic Gaussian fit to the WMAP-Q1 beam versus
multipole corresponding to the different order of the perturbation
expansion of a window function for a non-circular beam.
Note that the effect kicks in at $l\bar\sigma \sim 1$. }
\label{fig:beam}
\end{figure}

In order to find an expression for window function in terms of the
$\B_{lm}$ and $B_l$, we follow the approach in~\cite{TR:2001}. The
beam transforms for an arbitrary pointing direction $\q$ may be
expressed as,
\begin{equation}
b_{lm}(\q) \ = \ \sum_{m^\p=-l}^l b_{lm^\p}(\mathbf{\hat{z}})
D_{mm^\p}^l(\q,\rho(\q)), \label{eq:TR}
\end{equation}
where $D_{mm^\p}^l(\q,\rho) \equiv D_{mm^\p}^l(\phi,\theta,\rho)$ are
the Wigner-$D$ functions given in terms of the Euler angles describing the
rotation that carries the pointing direction $\q$ to
$\mathbf{\hat{z}}$-axis, as illustrated in Figure~\ref{fig:sphere}.
The third angle $\rho(\q)$ measures the angle by which the beam has
rotated about the new pointing direction, when the pointing direction
moves from $\mathbf{\hat{z}}$ to $\q$~\footnote{Hereafter, for brevity
of notation, absence of the pointing direction argument to $b_{lm}$ or
$\B_{lm}$ will imply a beam pointed along the $\mathbf{\hat{z}}$
axis.}. Inserting the spherical transform of the beam in
eq.~(\ref{eq:TR}) into eq.~(\ref{eq:defW}) we can write the
window function as
\begin{eqnarray}
W_l(\q_1,\q_2) &=& \frac{4\pi}{2l+1} \sum_{m=-l}^{l} b_{lm}^*(\q_1)
b_{lm}(\q_2)\\ &=& B_l^2 \sum_{m_1=-l}^{l} \sum_{m_2=-l}^{l}
\B_{lm_1}^* \B_{lm_2} \sum_{m=-l}^l D_{mm_1}^{l*}(\q_1,\rho(\q_1))
D_{mm_2}^{l}(\q_2,\rho(\q_2)) \label{eq:WD}
\end{eqnarray}
solely in terms of the circular component of the beam function $B_l$
and non-circular parts encoded in the BDP's, $\B_{lm}$. As pointed out
in ~\cite{TR:2001}, the window function expressed in the form of
eq.~(\ref{eq:WD}) has an obvious expansion in perturbation series in
$\B_{lm}$ retaining only the lowest values of $|m_1|$ and $|m_2|$. In
this paper, we adopt this perturbation approach to evaluate the
leading order correction to power spectrum estimation arising due to
mild deviations of the beam from circular symmetry.

For numerical evaluation it is advantageous to use the summation
formula of Wigner-$D$ to combine the product of the two Wigner-$D$
functions in eq.~(\ref{eq:WD}) into a single one as ~\cite{TR:2001}

\begin{equation}
   W_l(\q_1,\q_2) = {4\pi \over 2l +
   1} \sum^l_{m^\prime = - l} \, \sum^l_{m^{\prime\prime} =
   - l} \left[b_{l m^\prime}(\mathbf{\hat z})\right]^* b_{l
   m^{\prime\prime}}(\mathbf{\hat z}) D^l_{m^\prime
   m^{\prime\prime}}(\alpha -\rho_1,\, \gamma,\, \beta+\rho_2),
   \label{wigwine}
\end{equation}
where
\begin{eqnarray}
   {\rm cos} \gamma & = & \q_1\cdot\q_2 \nonumber \\
   {\rm cot} \alpha & = & - {\rm cos}\theta_1\, {\rm cot}  (\phi_1 - \phi_2) +  
   {\rm sin}\theta_1\, {\rm cot}\theta_2\, {\rm csc} (\phi_1 - \phi_2)
\label{wigwine_aux} \\
   {\rm cot} \beta & = & - {\rm cos}\theta_2\, {\rm cot}  (\phi_1 - \phi_2) +  
   {\rm cot}\theta_1\, {\rm sin}\theta_2\, {\rm csc} (\phi_1 - \phi_2) .
   \nonumber    
\end{eqnarray}

For large values of $l$ it is computationally expensive to evaluate
the entire $m^\prime$ and $m^{\prime\prime}$ sum in
eq.~(\ref{wigwine}).  However, for a smooth, mildly non-circular beam
function, restricting the summation to a few low values of
$m^\prime$ and $m^{\prime\prime}$ results in a good approximation.
The leading order terms in the perturbation~\cite{TR:2001}
\begin{eqnarray}
   W_l(\q_1,\, \q_2) &=& {4\pi \over 2l + 1} 
    \bigg[ \left[b_{l 0} (\mathbf{\hat z})\right]^2
                d^l_{00}(\gamma) + 2 b_{l 0} (\mathbf{\hat z}) 
                b_{l 2} (\mathbf{\hat z})
                \left\{ {\rm cos}(2(\alpha-\rho_1) ) 
+ {\rm cos}(2(\beta+\rho_2)) \right\}
                d^l_{02}(\gamma) \nonumber \\
   & {} & + 2 \left[b_{l 2} (\mathbf{\hat z})\right]^2
              \left[ {\rm cos}(2(\alpha + \beta + \rho_2-\rho_1)) d^l_{22}(\gamma)
              + (-1)^{-l} {\rm cos}(2(\alpha - \beta -\rho_1-\rho_2))) 
              d^l_{22}(\pi - \gamma)\right] \nonumber \\
   & {} & + 2 b_{l 0} (\mathbf{\hat z}) b_{l 4} (\mathbf{\hat z})
              \left\{ {\rm cos}(4(\alpha-\rho_1)) + {\rm cos}(4(\beta+\rho_2)
) \right\}
              d^l_{04}(\gamma) + \cdots \bigg].
   \label{wigwin40}
\end{eqnarray}
can be readily evaluated using recurrence relations similar to that of
Legendre function. In the above we have restricted to the common
situation of beam functions with reflection symmetry ($\B_{lm}$ are
real and $\B_{lm}=0$ for odd $m$) such as the elliptic Gaussian beam
described next.

An elliptic Gaussian beam profile, pointed along the $\mathbf{\hat
z}$-axis is expressed in terms of the spherical polar
coordinates about the pointing direction as follows~\cite{TR:2001}
\begin{equation} 
B(\mathbf{\hat{z}},\q) \ = \ \frac{1}{2\pi\sigma_1\sigma_2}\exp \left[ -\frac{\theta^2}{2\sigma^2(\phi)} \right],
\end{equation}
where the ``beam-width" $\sigma(\phi) \ \equiv \
[\sigma_1^2/(1+\epsilon \sin^2\phi)]^{1/2}$ and the ``non-circularity
parameter" $\epsilon \ \equiv \ (\sigma_1^2/\sigma_2^2 - 1)$ are given
in terms of $\sigma_1$ and $\sigma_2$ -- the Gaussian widths along the
semi-major and semi-minor axis, respectively. However, we characterize
an elliptical beam using two different parameters: eccentricity $e
\equiv \sqrt{1-\sigma_2^2/\sigma_1^2}$ and the size parameter
$\theta_{1/2}$, the FWHM of a circular beam of equal ``area"\footnote{
By ``area" we mean the area enclosed by the curve whose each point
corresponds to the Half Maximum of the Gaussian profile. We can show
that, $\theta_{1/2}(\mbox{in degrees}) = (180/\pi)\sqrt{8\ln{2}}
\,\bar\sigma$, where $\bar\sigma^2 \equiv \sigma_1\sigma_2$ is
proportional to the area of the beam.}.

For elliptical Gaussian beams the spherical harmonic transform is
available in the closed analytical form
\begin{equation}
b_{lm} \ = \ \left[ \frac{2l+1}{4\pi} \, \frac{(l+m)!}{(l-m)!}
\right]^{\frac{1}{2}} (l+1/2)^{-m} I_{m/2} \left[ \frac{(l+1/2)^2
\sigma_1^2 e^2}{4} \right]\,\exp \left[ -\frac{(l+1/2)^2
\sigma_1^2}{2} \left\{ 1 - \frac{e^2}{2} \right\} \right]\,,
\label{eq:blm:elliptical}
\end{equation}
where $I_\nu(x)$ is the modified Bessel
function~\cite{TR:2001,Challinor}.  Note, in the above equation we
have used eccentricity $e$ instead of the non-circularity parameter
$\epsilon = e^2/(1-e^2)$ used in \cite{TR:2001}. (Please see
Table~\ref{tab:2} for the various definitions and characterizations of
elliptical beams.)

\begin{table}
\caption{In literature, the elliptical beams have been described by
several parameters which can all be expressed in terms of the
Gaussian widths along the semi-major ($\sigma_1$) and the
semi-minor ($\sigma_2$) axes of the ellipse. We have used these
parameters at several places in the paper.}
\begin{center}
\begin{tabular}{lcc}
\hline
Parameter                 & Symbol          & Expression \\ \hline \hline
Eccentricity              & $e$             & $\sqrt{1 - \frac{\sigma_2^2}{\sigma_1^2}}$ \\
Non-Circularity Parameter & $\epsilon$      & $\frac{\sigma_1^2}{\sigma_2^2} - 1$ \\
Ellipticity               & $\bar\epsilon$  & $ 1- \frac{\sigma_2}{\sigma_1}$\\
\hline
\label{tab:2}
\end{tabular}
\end{center}
\end{table}

Fig~\ref{fig:beam} shows one of the WMAP beams as an example of a
distinctly non-circular beam (see iso-contours in the left panel) that
can be efficiently handled by the leading order term in the
perturbation approach (see the right panel).  Details of the exercise
of fitting elliptical Gaussian beam profile to the WMAP beam maps is
given in appendix~\ref{wmapbeamfit}.

%%%%%%%%%%%%%%%%%%%%%%%%%%%%%%%%%%%%%%%%%%%%%%%%%%%%%%%%%%%%%%%%%%%%%%%%%%%

\section{Bias Matrix }
\label{secBias}

Given the observed temperature fluctuations $\delTobs(\q)$, a naive estimator
for the angular power spectrum based on eq.~(\ref{defcl}) is given by
\begin{equation} 
  \widetilde{\mathcal{C}}_l \equiv \ \frac{l(l+1)}{2\pi} \frac{1}{2l+1}
\sum_{m=-l}^l |\tilde a_{lm}|^2 ,
\label{pCl}
\end{equation}
where 
\begin{equation}
\tilde a_{lm} \ \equiv \ \int d\Omega_\q \, Y_{lm}^*(\q) \, \delTobs(\q)
\,w(\q)
\end{equation}
are the coefficients of the spherical harmonic transform of the CMB
anisotropy map~\cite{yu_peeb69,peeb73}. The weight function $w(\q)$
accounts for non-uniform/incomplete sky coverage and also provides a
handle to weigh the data `optimally'.  Without the
inconsequential $l(l+1)$ scaling, this naive estimator is
referred to as the pseudo-$C_l$ in recent
literature~\cite{wan_hiv03}. The `pseudo' refers to fact that the
estimated $C_l$ is biased. Moreover, this is a sub-optimal estimator of
the power spectrum. This naive power spectrum estimate has to be
corrected for observational effects such as the instrumental noise
contribution, beam resolution, incomplete/non-uniform sky
coverage. Nevertheless, the pseudo-$C_l$ method is a computationally
fast and economical approach and is currently a method of choice for
the recent large CMB anisotropy datasets (at least for large $l$
within the hybrid schemes~\cite{efs04}).

Faced with the computational challenges of large data sets, an
approach that has been adopted is to compute the pseudo-$C_l$'s from
the CMB observations and then correct for the observational
effects. The true $C_l$ spectrum is linearly related
\begin{equation} 
 \langle\tilde {\mathcal{C}}_l\rangle = \sum_{\lp} A_{l\lp}\,
 {\mathcal{C}}_\lp
\end{equation}
to the pseudo-$C_l$ through a \emph{bias} matrix $A_{l\lp}$. Similar
bias matrices arising due to the effect of non-uniform sky coverage,
instrumental noise have been studied~\cite{wan_hiv03,master}. In this
paper, we compute the $A_{l\lp}$ for non-circular beam and give
explicit analytical results for the leading order terms for non
rotating beams.

The pseudo-$C_l$ estimator in eq.~(\ref{pCl}) can be expressed as
\begin{equation} 
\widetilde{\mathcal{C}}_l \ \equiv \ \frac{l(l+1)}{8\pi^2} \int
d\Omega_{\q_1} \int d\Omega_{\q_2}\, w(\q_1) w(\q_2)\,\delTobs(\q_1)
\delTobs(\q_2) P_l(\q_1\cdot\q_2).\label{eq:clobs}
\end{equation}

The ensemble expectation value of the pseudo-$C_l$ power spectrum
estimator is 
\begin{eqnarray}
\cclobs &=& \frac{l(l+1)}{8 \pi^2} \int d\Omega_{\q_1} \int
d\Omega_{\q_2}\, w(\q_1) w(\q_2) \, \times \nonumber\\
&&\sum_\lp \frac{2\lp+1}{2\lp(\lp+1)} \mathcal{C}_\lp
P_l(\q_1 \cdot \q_2) \int d\Omega_\q \int d\Omega_\qp \, B(\q_1, \q)
B(\q_2, \qp) P_\lp(\q \cdot \qp).
\end{eqnarray}

Recalling the definition of a window function in eq.~(\ref{windef}), the
most general form of the bias matrix
\begin{equation} 
A_{l\lp} \ = \ \frac{2\lp+1}{16\pi^2} \frac{l(l+1)}{\lp(\lp+1)}
\int d\Omega_{\q_1} \int d\Omega_{\q_2}\, w(\q_1) w(\q_2)\, P_l(\q_1
\cdot \q_2) W_\lp(\q_1,\q_2). \label{eq:defA}
\end{equation}

Using the expression for the window function for a non circular beam
in eq.~(\ref{eq:WD}) the bias matrix can be written as
\begin{equation} 
A_{l l^\prime} = \ \frac{B_\lp^2}{4\pi} \frac{(2l^\prime+1)}{(2l+1)}
\frac{l(l+1)}{l^\prime(l^\prime+1)} \times \sum_{n=-l}^{l}
\sum_{m=-\lp}^{\lp} \left|\sum_{m^\p=-\lp}^{\lp} \B_{\lp m^\p} \int
d\Omega_\q\, Y_{ln}^*(\q) \, D_{mm^\p}^{\lp}(\q,\rho(\q)) \,w(\q)
\right|^2. \label{eq:a}
\end{equation}
The above expressions in eq.~(\ref{eq:defA}) and eq.~(\ref{eq:a}) are
valid for a completely general non-circular beam with an arbitrary
orientation at each point. The scan pattern of the CMB experiment and
relative orientation of the beam along it is encoded in the function
$\rho(\q)$. The weight $w(\q)$ can account for non-uniform sky
coverage.  Analytical progress can be made when $w(\q)\equiv
w(\theta)$ and $\rho(\q)=\rho(\theta)$ are fixed along a given
declination, but we do not discuss further it here.  When the beam
transform, weight function and the scan pattern are specified, the
bias matrix can be evaluated numerically using eq.~(\ref{eq:a}).
However, for mild deviations from circularity, the above expression
also points to a perturbation expansion in the small beam distortion
parameters, $\B_{lm}$.

For obtaining fully analytical results, we set the weight function
$w(\q)=1$, corresponding to a full, uniform sky coverage and also
limit attention to scans with `non-rotating' beams where $\rho(\q)=0$.
This is presented in the next subsections

\subsection{Circular Symmetric Beam}

We first consider eq.~(\ref{eq:a}) for the simpler and well studied
case of a circular beam. For clarity of presentation, we limit our
discussion full, uniform sky coverage ($w(\q)=1$). Results for
non-uniform coverage with a circular beam are available in the
literature~\cite{wan_hiv03,master,efs04}.

Using the expression for the window function for circular beam
eq.~(\ref{isowine}) into the expression for the bias in
eq.~(\ref{eq:defA}) we recover
\begin{equation}
A_{l\lp} \ = \ B_{l}^2 \, \delta_{l\lp} \ \ \Rightarrow \ \cclobs \ =
\ B_{l}^2 \, \mathcal{C}_l. \label{eq:Cliso}
\end{equation}

For a full sky measurement with a circular beam, the bias matrix is
diagonal implying that there is no mixing of power between different
multipoles. The true expectation value of the power spectrum can be
obtained by dividing the pseudo-$C_l$ estimator by the isotropic beam
transform $B_{l}^2$.

Next we account for the noise contribution and recover the well known
result for a full sky observation.  The pixel noise $n(\q)$ adds to
the observed temperature, so that the resultant observed temperature

\begin{equation} 
\delTobs^\p(\q) \ = \ \delTobs(\q) + n(\q)
\end{equation}
and we can readily obtain
\begin{equation} 
\cclobsp\ = \ \cclobs \ + \ \mathcal{C}_l^N \ = \ B_{l}^2 \,
\mathcal{C}_l \ + \ \mathcal{C}_l^N, \label{eq:eclp}
\end{equation}
where $\mathcal{C}_l^N $ is the angular power spectrum of the noise
$n(\q)$ is a well determined quantity.  The unbiased estimator for
$\mathcal{C}_l$ obtained is 
\begin{equation} 
\club \ = \ B_{l}^{-2} \, \left( \widetilde{\mathcal{C}}_l^\p \ - \
\mathcal{C}_l^N \right).
\label{clcirub}
\end{equation}

%%%%%%%%%%%%%%%%%%%%%%%%%

\subsection{Non-circular Beam}

We obtain analytic results for the bias matrix for a full sky
observation ($w(\q)=1$) with a non-circular beam that `does not
rotate'.  The phrase ``non-rotating" means that the orientation
of the non-circular beam does not
rotate about its axis (the pointing direction) while the pointing
direction scans the sky implying that
\begin{equation} 
\rho(\q) \ = \ 0.
\end{equation}
For non-rotating beam, the calculation of the bias is completely analytically
tractable. The integral in the expression for the bias in eq.~(\ref{eq:a})
is given by
\begin{equation} 
\int Y_{ln}^*(\mathbf{\hat{q}}) \,
D_{mm^\prime}^{\lp}(\mathbf{\hat{q}},0) \, d\Omega_\q \ = \
\sqrt{(2l+1)\pi} \ I_{mm^\p}^{l\lp} \, \delta_{mn},
\label{eq:intYD}
\end{equation}
where
\begin{equation} 
I_{mm^\p}^{l\lp} \ \equiv \ \int_{-1}^{1} d_{m0}^l(\theta) \, d_{mm^\p}^{\lp}(\theta) \, d\cos\theta,
\end{equation} \label{eq:defI}
and $d_{mm^\p}^l(\theta)$ are Wigner-$d$ functions related to Wigner-$D$ functions
\begin{equation} 
D_{mm^\p}^l(\q,\rho) \ = \ e^{-im\phi} \, d_{mm^\p}^l(\theta)
e^{-im^\p\rho}.
\end{equation}
The analytic simplicity arises from the fact that for $\rho(\q)=0$, the
Wigner-$D$ function reduces to spherical harmonic function
\begin{equation} 
Y_{lm}(\q) \ = \ \sqrt{\frac{2l+1}{4\pi}} \, D_{m0}^{l}(\q,0)\,.
\end{equation}
In deriving the above have used the orthogonality of the phases
$\int_0^{2\pi} \, e^{-i(m-n)\phi} d\phi \ = \ 2\pi \delta_{mn}$.

Substituting the expression for the integral eq.~(\ref{eq:intYD}) into
the expression for the bias in eq.~(\ref{eq:a}), we obtain
\begin{equation} 
A_{l\lp} \ = \ B_\lp^2 \left(\frac{2\lp+1}{4}\right)\frac{l(l+1)}{\lp(\lp+1)} \sum_{m=-L}^{L} \left|\sum_{m^\p=-\lp}^{\lp} \B_{\lp m^\p} \ I_{mm^\p}^{l\lp} \right|^2,
\end{equation}
where $L\equiv{\rm min}\{l,\lp\}$ is the smaller between $l$ and $\lp$.

Further analytical progress is possible for smooth beam with mild
deviations from circular symmetry through a perturbation in terms of
the small beam distortion parameters, $\B_{lm}$. We calculate
the exact analytic expression for the leading order effect.  Assuming
a beam with reflection symmetry where $\B_{lm}$ are zero for odd $m$,
the leading order effect comes at the second order, namely,
$\B_{l2}\B_{l0}$ (see eq.~(\ref{wigwin40})). Neglecting, $\B_{lm}$ for
$|m|>2$, we obtain
\begin{equation} 
A_{l\lp} \ = \ B_\lp^2 \left(\frac{2\lp+1}{4}\right)
\frac{l(l+1)}{\lp(\lp+1)} \sum_{m=-L}^{L} \left[ I_{m0}^{l\lp} \, + \,
\B_{\lp 2} (I_{m2}^{l\lp} + I_{m-2}^{l\lp}) \right]^2.
\end{equation}
Next we obtain analytical expression for the two integrals,
$I_{m0}^{l\lp}$ and $I_{m2}^{l\lp}+I_{m-2}^{l\lp}$. The first one can
be found in standard texts (e.g. \cite{VMK}) given as
\begin{equation}
I_{m0}^{l\lp} \ \equiv \ \int_{-1}^{1} \, d_{m0}^l(\theta)
d_{m0}^{\lp}(\theta) d\cos\theta \ = \ \frac{2}{2l + 1}\delta_{l\lp}.
\end{equation}
For $m=0$, writing $d_{00}^l(\theta)$ and $d_{02}^l(\theta)$ in terms
of $P_l(\cos\theta)$ and its first derivative $P_l^\p(\cos\theta)$ we have shown in
Appendix-\ref{Appint} that for odd values of $l+l^\prime$,
$I_{02}^{l\lp}+I_{0-2}^{l\lp}=0$. For even values of $l+l^\prime$,

\begin{equation} \label{eq:Im0}
I_{02}^{l\lp}+I_{0-2}^{l\lp} \ = \ \left\{
\begin{array}{ll}
8/\kappa                 & \mbox{if $l < \lp$} \\ \\
0                        & \mbox{if $l > \lp$,} \\ \\
-(4l/\kappa)(l-1)/(2l+1) & \mbox{if $l = \lp$} \\
\end{array} \right.
\end{equation}
where $\kappa \ \equiv \ \sqrt{(\lp-1)\lp(\lp+1)(\lp+2)}$.

To evaluate $I_{m2}^{l\lp}+I_{m-2}^{l\lp}$ for non-zero $m$ we expand
$d_{m\pm2}^\lp(\theta)$ in terms of $d_{m0}^{l^{\p\p}}(\theta)$ using
a recurrence relation of the Wigner-$D$ functions (where $l^{\p\p}$ takes
integer values between $\lp-2$ to $\lp+2$) . The details are given in
Appendix-\ref{Appint}.  We obtain that
$I_{m2}^{l\lp}+I_{m-2}^{l\lp}=0$ for odd $l+l^\prime$. For even values
of $l+l^\prime$, if $L\equiv{\rm min}\{l,\lp\}\ge |m|>0$,
\begin{equation} 
I_{m2}^{l\lp} + I_{m-2}^{l\lp} \ = \ \left\{ \begin{array}{ll}
(4/\kappa)(|m|+1) \sqrt{\frac{(l+|m|)!(\lp-|m|)!}{(l-|m|)!(\lp+|m|)!}}
& \mbox{if $l < \lp$}\\ \\
(4/\kappa)(|m|-1)\sqrt{\frac{(l-|m|)!(\lp+|m|)!}{(l+|m|)!(\lp-|m|)!}}
& \mbox{if $l > l^\prime$}\\ \\ (4/\kappa) [|m| \, - \,
(l^2+l+1)/(2l+1)] & \mbox{if $l = \lp$.}
\end{array} \right. \label{eq:Im2}
\end{equation}

\begin{figure}
\centering
\includegraphics[width=0.7\textwidth,angle=0]{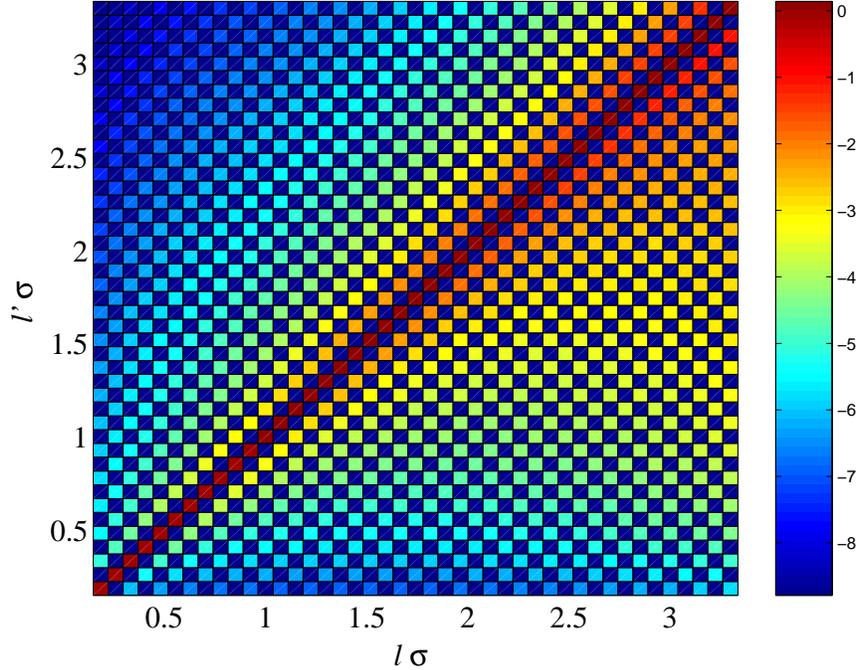}
\caption{The log of normalized bias matrix $A_{ll'}/(B_lB_\lp)$ is plotted for an
elliptical beam of eccentricity $e=0.6$ and mean beam-width
$\bar\sigma = 0.074$. The normalization is carried out so that the
effect of non-circularity on the bias matrix can be easily compared to
that for circular beams.  Beam rotation and cut-sky effects have not
been considered in this figure.  One notices that the off-diagonal
elements of the bias matrix take significant values for $l \bar\sigma \geq
1$.}
\label{fig:biasmat}
\end{figure}

The bias matrix including the leading order beam distortion (for
non-rotating, reflection symmetric beams) can be summarized as

\noindent
$\bullet$ For odd values of $l+ \lp$,
\begin{equation}
A_{l l^\prime} \ = \ 0.
\end{equation}

\noindent
$\bullet$ For even values of  $l + l^\prime$,
\begin{equation}
A_{l\lp} \ = \ \left\{ \begin{array}{ll}

(B_\lp\B_{\lp 2})^2 \left( \frac{8 \,
l(l+1)(2l^\prime+1)}{l^{\prime2}(l^\prime+1)^2(l^\prime-1)(l^\prime+2)}
\right) \left[ 2 \, + \, \sum_{m=1}^l
\frac{(l+m)!(l^{\prime}-m)!}{(l-m)!(l^{\prime}+m)!} (m+1)^2 \right] &
\mbox{if $l < l^\prime$} \\ \\

(B_\lp\B_{\lp 2})^2 \left( \frac{8 \,
l(l+1)(2l^\prime+1)}{l^{\prime2}(l^\prime+1)^2(l^\prime-1)(l^\prime+2)}
\right) \left[ \sum_{m=1}^{l^\prime}
\frac{(l-m)!(l^{\prime}+m)!}{(l+m)!(l^{\prime}-m)!} (m-1)^2 \right] &
\mbox{if $l > l^\prime$} \\ \\

\frac{B_{l}^2}{2l+1} \left[ \left\{1 - 2 \B_{l2} \,
\sqrt{\frac{l(l-1)}{(l+1)(l+2)}} \right\}^2 + 2 \sum_{m=1}^{l} \left\{
1 - 2 \B_{l2} \, \frac{(l^2+l+1)-(2l+1)m}{\sqrt{(l-1)l(l+1)(l+2)}}
\right\}^2 \right] & \mbox{if $l = l^\prime$}.
\end{array} \right.
\end{equation}

The non-zero off-diagonal terms in the bias matrix $A_{l\lp}$ imply
that the non-circular beam mixes the contribution of different
multipoles from the actual power spectrum in the observed power
spectrum. Off-diagonal elements in $A_{l\lp}$ that arise from
non-uniform/incomplete sky coverage have been studied earlier and are
routinely accounted for in CMB experiments. Non-circular beam is yet
another source of off-diagonal terms in the bias matrix and should be
similarly taken into account. In general, CMB experiments have both
non-circular beams and non-uniform/incomplete sky coverage that could
lead to interesting features in $A_{l\lp}$.

Although the analytical result is limited to mildly non-circular and
non-rotating beam functions, it does bring to light certain generic
features of the effect of non-circular beam functions. To be specific,
we compute the elements $A_{l\lp}$ for non-rotating \emph{elliptic
Gaussian} beams (see appendix~\ref{wmapbeamfit}). The non-circularity
of these beams is characterized by their eccentricity
$e=\sqrt{1-\sigma_2^2 /\sigma_1^2}$, where $\sigma_1$ and $\sigma_2$
are the $1\sigma$ beam-widths along major and minor axes of the beam
(see table~\ref{tab:2}). Many experiments have characterized their
beams in terms of an elliptic Gaussian fit
(e.g.,\cite{pyV,whu01,fos02}).  A convenient advantage of elliptical
beams is that the beam transform $b_{lm}$ (and obviously, the beam
distortion parameters, $\B_{lm}$) can be expressed in a closed
analytical form. The results expressed in terms of $l\bar\sigma$ are
broadly independent of the average beam-size~\cite{TR:2001}.

\begin{figure}[h]
\centering
\includegraphics[width=0.45\textwidth]{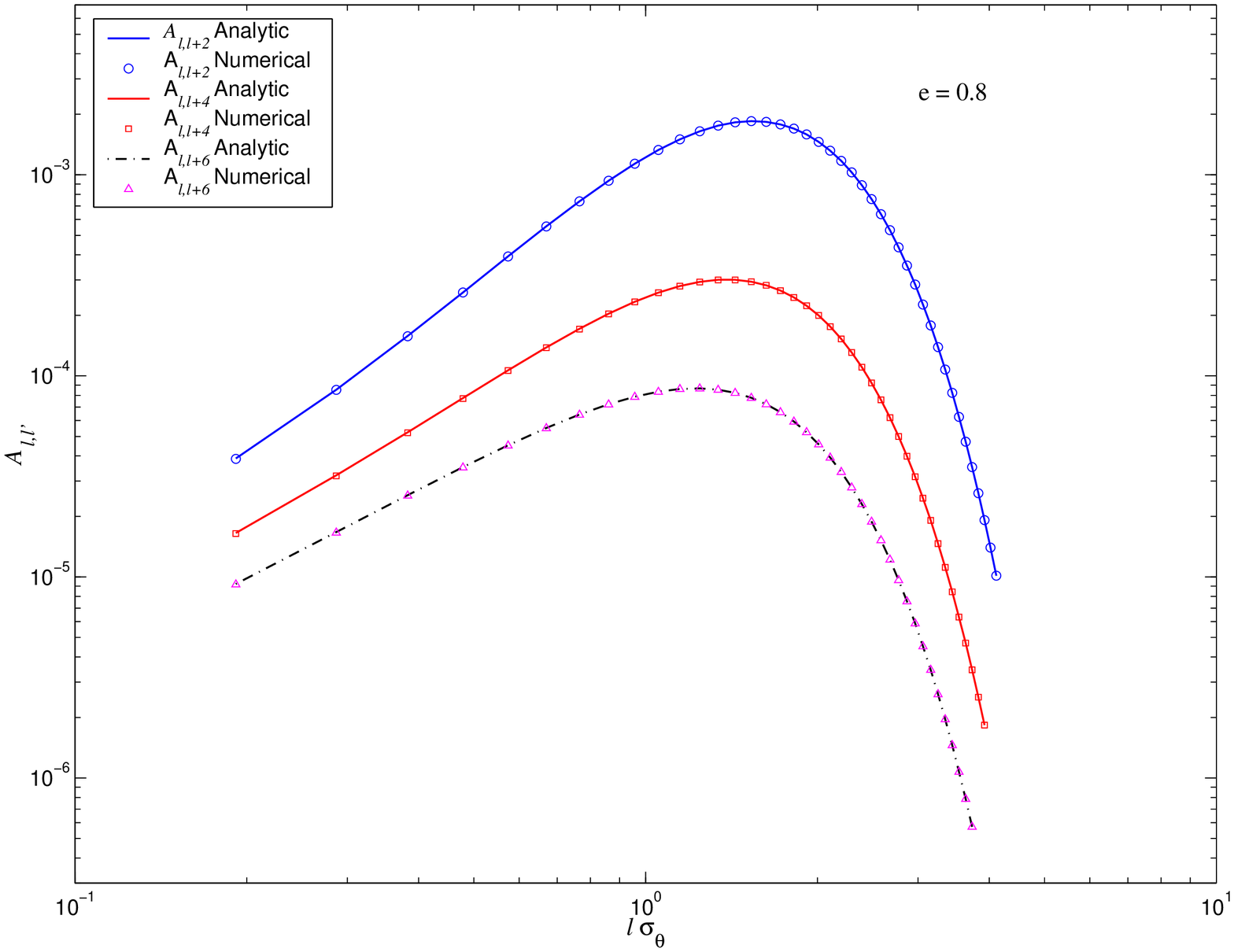}
\includegraphics[width=0.45\textwidth]{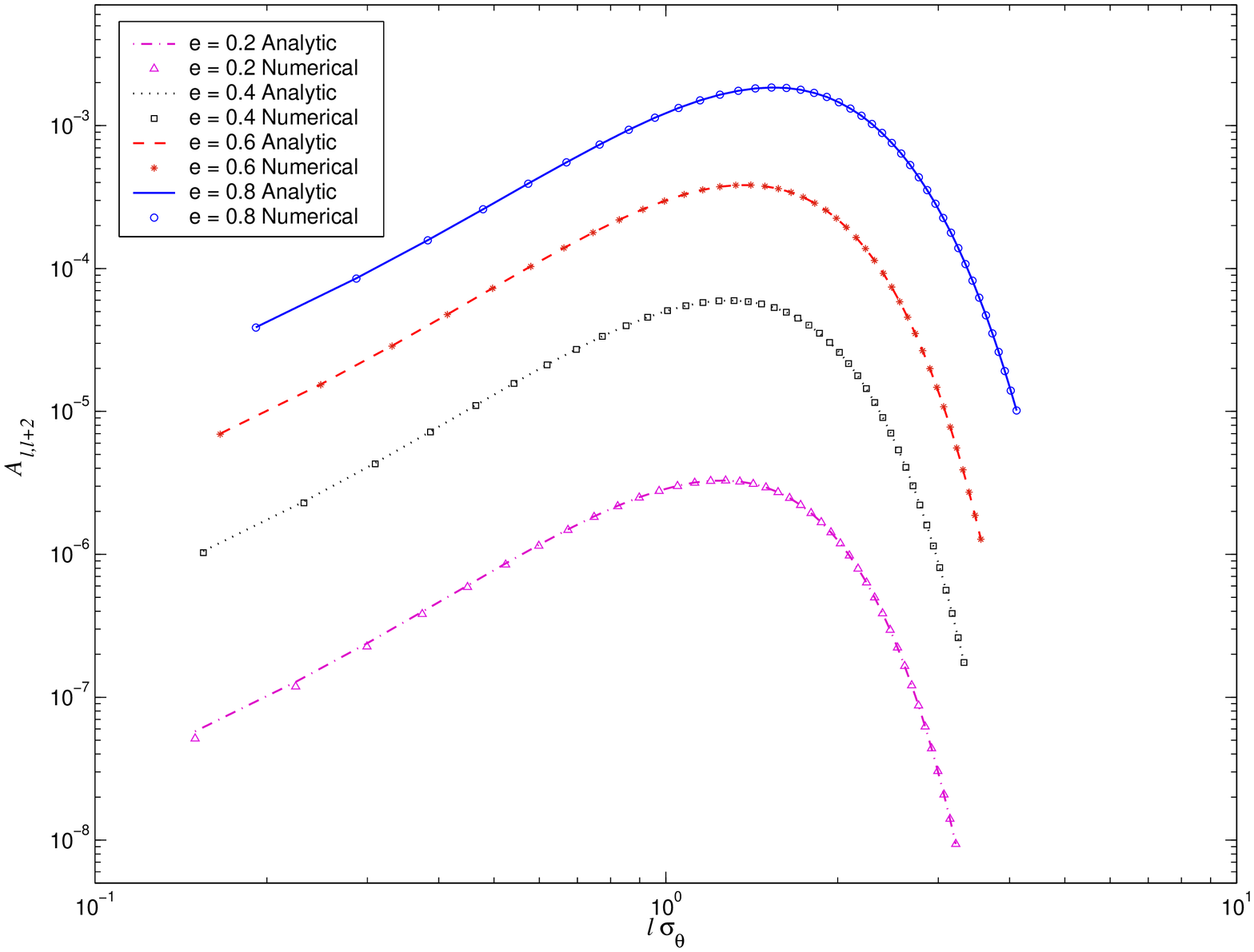}
\caption{Elements of the bias matrix $A_{l\lp}$ are plotted in this
figure as a function of multipole $(l)$. The bias matrix relates the
observed $\mathcal{C}_l$s to their true values. When non-circular beams are used
in CMB experiments, the bias matrix can be shown to be non-diagonal,
thus implying mixing of power between multipoles. On the left panel,
we plot $A_{l\lp}$ for $\lp-l = 2,4,6$. It is evident that the effect
decreases as we move away from the diagonal and that it kicks in at $l
\bar\sigma \sim 1$, for a beam of eccentricity $e=0.8$. For the figure
in the right panel, we plot $A_{l\, l+2}$ for several beams of the
same size but different eccentricities. Clearly, the effect also
depends strongly on the non-circularity of the beam.}
\label{fig:Allp}
\end{figure}

Fig.~\ref{fig:biasmat} shows a density plot of the normalized bias
matrix $A_{l\lp}/(B_lB_\lp)$ for a non-rotating elliptical beam.  The
plot illustrates the importance of off-diagonal terms that arise due
to the non-circular beam relative to the diagonal terms. The absence
of coupling between multipoles separated by odd integers is
evident. Also evident is the fall off as one moves away from the
diagonal. The left panel of Fig.~\ref{fig:Allp} shows that the
off-diagonal elements of $A_{l\lp}$ are important at $l\bar\sigma\sim
1$. The results are qualitatively independent of the average beam size
$\bar\sigma$. The right panel Fig.~\ref{fig:Allp} shows the strong
dependence of the dominant off-diagonal element $A_{l\, l+2}$ on the
eccentricity of the beam.

The analytical results and numerical computations using
eq.~(\ref{eq:defA}) were compared.  The numerical and analytical
results match perfectly as shown in Figure~\ref{fig:Allp}. Numerical
computation involves the pixelized sky and the algorithm must ensure
that this does not introduce spurious effects. We verify that
$A_{l\lp}$ has numerically negligible off-diagonal elements when the
beam is circularly symmetric. The numerical computation for non-circular
beam are verified to be robust to the pixelization of the sky.

\begin{figure}
\centering
\includegraphics[height=0.4\textwidth]{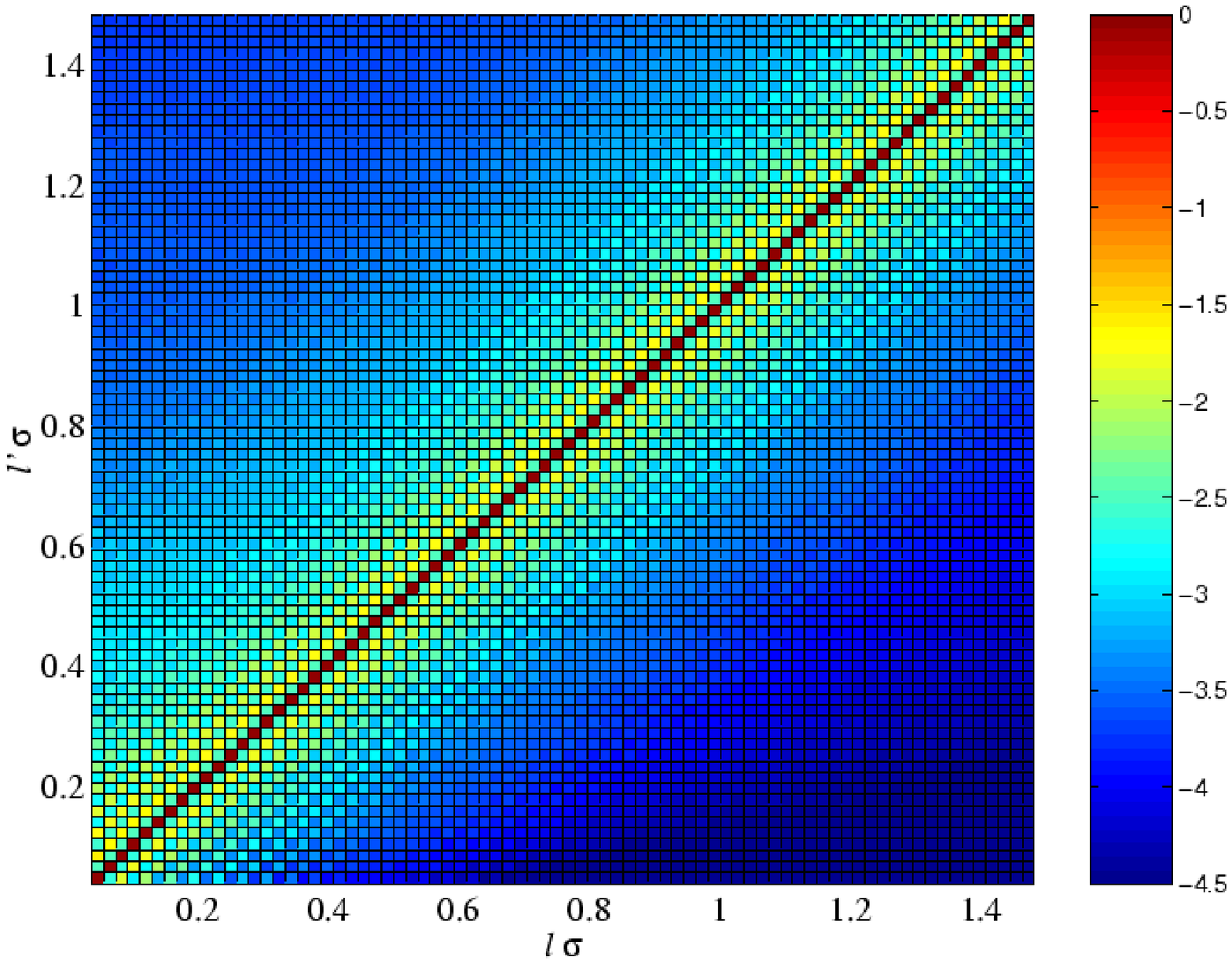}
\includegraphics[height=0.4\textwidth,width=0.43\textwidth]{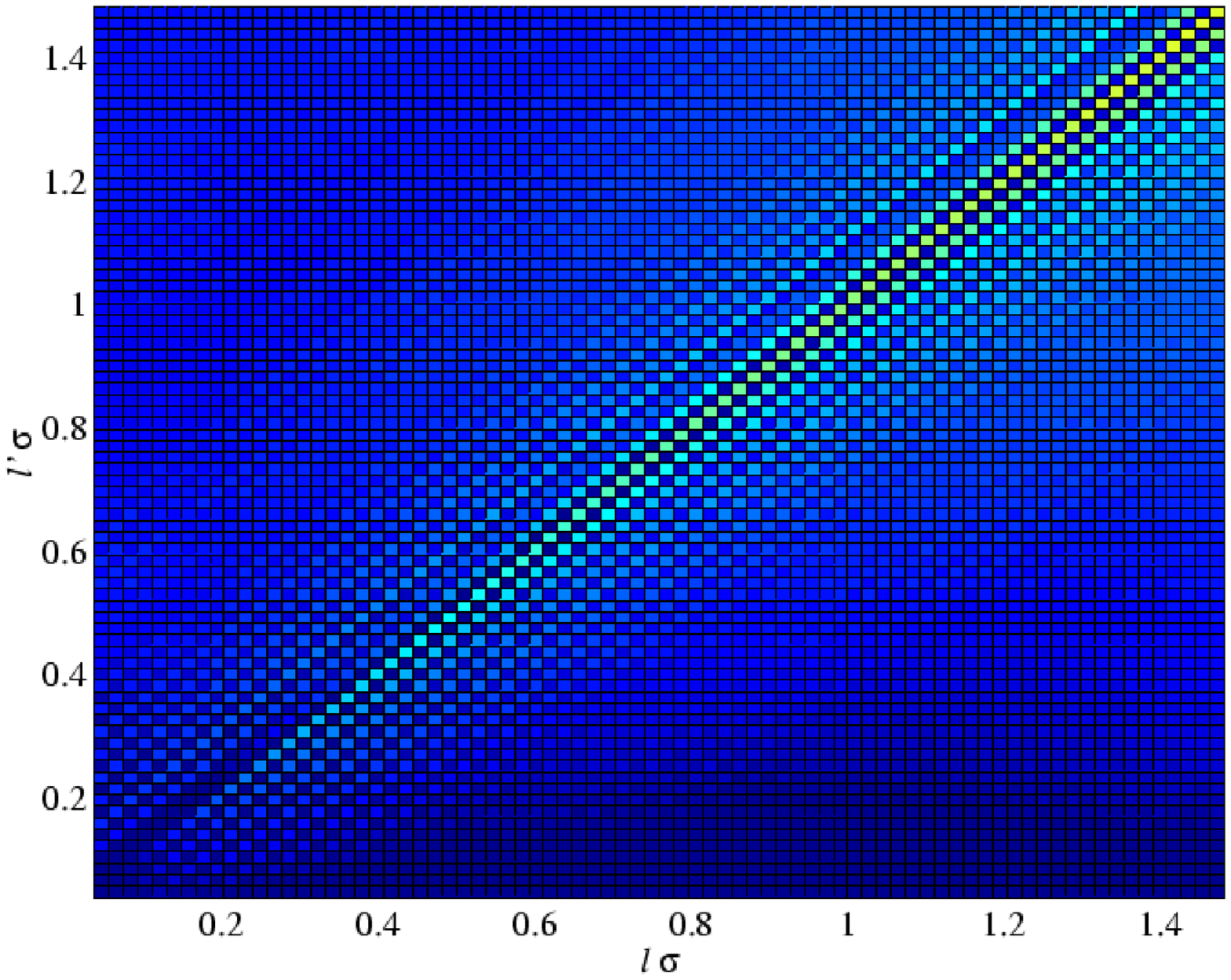}
\caption{ The normalized bias matrix elements (on log scale,
$\log[A_{l\lp}/(B_lB_\lp)]$) of a hypothetical experiment with a scan
pattern (eq.~\ref{eq:rotScheme}) corresponding to a rotating, non
circular beam ($e=0.6$) and non uniform sky coverage are studied. The
left panel shows the effect of non-uniform coverage alone (circular
beam approximation). The right panel isolates the additional effect
that arises due to the non-circularity of the beam and its
rotation. We note that significant off-diagonal elements arise at $l
\bar\sigma \geq 1$ from the non-circular beam comparable to that from
the non-uniform coverage. The non-uniform coverage corresponds to a
smoothed WMAP Kp2 galactic mask (smoothed from resolution of
$N_{side}=512$ to $64$). We use a sufficiently high resolution beam
with $\bar\sigma=0.018$ ($\theta_{1/2}=2.5^\circ$) to ensure that the
effects due to the galactic mask and the non-circular beam appear in
distinct regions of the multipole space.}
\label{fig:biasmatrix}
\end{figure}

Next we illustrate effect of beam-rotation and non-uniform sky
coverage for a \emph{hypothetical} experiment where $A_{l\lp}$ have
been computed numerically. The left panel of Fig.~\ref{fig:biasmatrix}
shows (in $log$ scale) the normalized bias matrix arising from a
$2.5^\circ$ circular beam including a non trivial $w(\q)$
in the form of a smoothed version of the galactic mask Kp2 of WMAP~\cite{Bennett:2003,LamdaBl}. The right panel of the figure
shows the \emph{extra} effect that a rotating non-circular beam
would introduce. We assume a simple `toy' beam rotation along
an equal declination scan strategy, where the beam continuously
`rotates' by $2\pi$ for every complete pass at a given declination
which implies the simple form
\begin{equation}
\rho(\q) \equiv \rho(\theta,\phi)  = \phi\,. \label{eq:rotScheme}
\end{equation}
The elements here have been computed numerically using
eq.~(\ref{eq:defA}) retaining the leading order terms in the
perturbation expansion of $W_l$ in eq.~(\ref{wigwin40}).  The
off-diagonal effects at low $l$ are dominated by the cut sky
effect. The off-diagonal element $l\bar\sigma\gsim 1$ arise solely due
to non-circular beam.  The numerical computation illustrates the
potentially large corrections that can arise due to non-circular beam
that `rotate' on the sky.  The numerical computations in this work
pave the way for introducing realistic scan-pattern, beam-rotation and
non-uniform sky coverage in a future extension to our work.

We summarize the following features of the bias matrix~:

\begin{enumerate}
\item There is no coupling between $\cclobs$ and $\mathcal{C}_\lp$
for odd values of $l + \lp $,
\item Coupling decreases as $|l - \lp|$ increases,
\item Coupling increases with eccentricity for fixed beam size, and
\item Size of the beam determines the multipole $l$ value for which
coupling will be maximum ($l\bar\sigma\sim 1$).\\
\end{enumerate}

\begin{figure}
   \centering
   \includegraphics[width=0.65\textwidth,angle=-90]{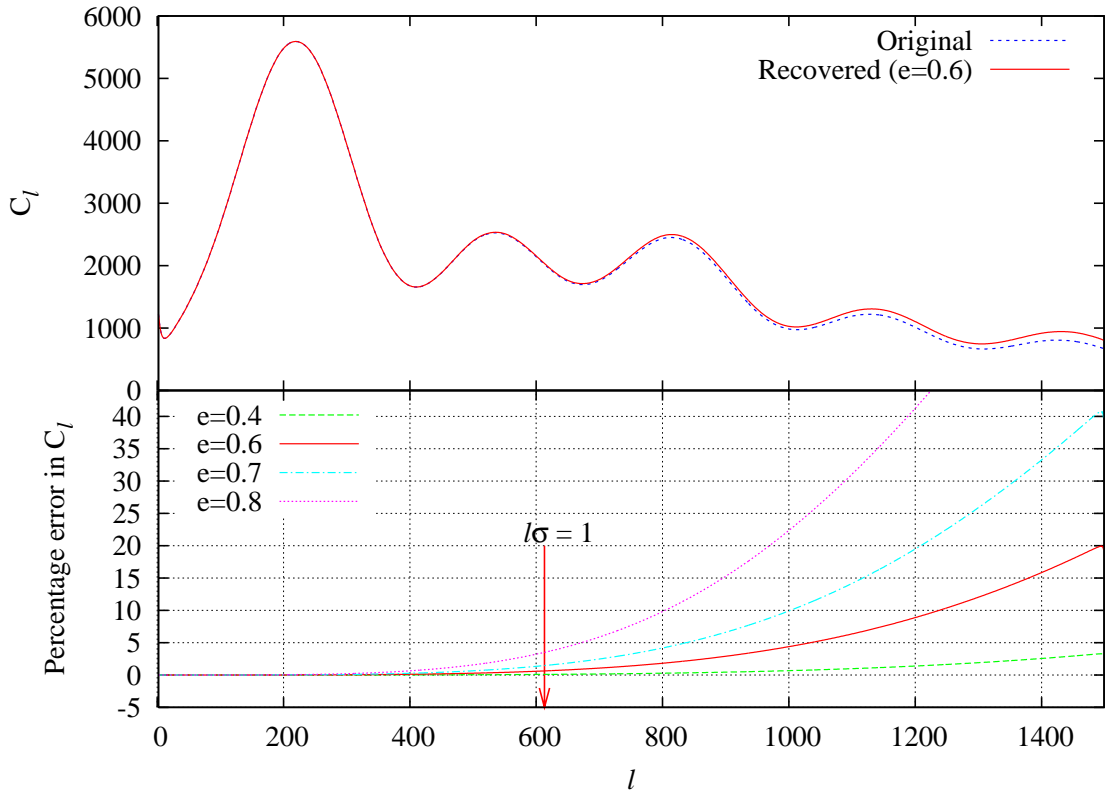}
\caption{The effect of non-circular beam is studied for CMB power
spectrum estimation by a CMB experiment with a WMAP-like non-circular
beam. For illustration, we consider the best fit (Power law) model to
be the (fiducial) true $\mathcal{C}_l$ of the Universe shown as the
solid line in the upper panel.  Let $\widetilde{\mathcal{C}}_l$ be the
power spectrum measured by using a elliptical, Gaussian beam with
eccentricity, $e=0.6$, and $\bar\sigma=0.0016$. The dashed line shows
the $\mathcal{C}_l$ that would be inferred by deconvolving
$\widetilde{\mathcal{C}}_l$ with a circular beam assumption with
beam-width, $\bar\sigma$. The lower panel plots the relative error in
the power spectrum recovered with a circular beam assumption for a
measurements made with a non-circular beam with $e=0.4$ to $0.8$.}
\label{fig:5}
\end{figure}

Figure~\ref{fig:5} roughly indicates the level and nature of the
effect of neglecting the non-circularity of the beam on CMB power
estimation (for the conservative case of non-rotating beams). Consider
the power spectrum $\widetilde{\mathcal{C}}_l = \sum_{\lp}\, A_{l\lp}
\mathcal{C}_\lp$ measured using a non-circular, elliptical
Gaussian beam of a given eccentricity, $e$ and average beam-width,
$\bar\sigma$. We compare the power spectrum obtained by deconvolving
$\widetilde{\mathcal{C}}_l$ with a circular, Gaussian beam of the
beam-width, $\bar\sigma$ with the true $\mathcal{C}_l$. The lower
panel shows that the error can be significant for multipole values
beyond the inverse beam-width even for modestly non-circular
comparable to the WMAP beam maps (Q-band) discussed in the
Appendix~\ref{wmapbeamfit}.

Finally, we construct the unbiased estimator for the angular power
spectrum. Invoking steps similar to the case of circular beams to
account for the instrumental noise, we obtain
\begin{equation} 
\cclobsp\ = \ \cclobs \ + \ \mathcal{C}_l^N \ = \ \sum_\lp A_{l\lp}
\mathcal{C}_\lp \ + \ \mathcal{C}_l^N.
\end{equation}
The unbiased estimator for the angular power spectrum is
\begin{equation} 
\club \ = \ \sum_\lp \, A_{l\lp}^{-1} \left(
\widetilde{\mathcal{C}}_\lp^\p \ - \ \mathcal{C}_\lp^N \right).
\label{clncub}
\end{equation}

%%%%%%%%%%%%%%%%%%%%%%%%%%%%%%%%%%%%%%%%%%%%%%%%%%%%%%%%%%%%%%%%%%%%%%%%%%%

\section{Error-Covariance Matrix}
\label{secCov}

The statistical error-covariance of the estimated angular power
spectrum is defined as
\begin{equation} 
\ccov \ \equiv \ \langle (\widetilde{\mathcal{C}}_l -
\cclobs)(\widetilde{\mathcal{C}}_\lp - \langle
\widetilde{\mathcal{C}}_\lp \rangle) \rangle.
\end{equation}
In an idealized, noise free, CMB experiment with infinite angular
resolution uniformly covering the full sky
\begin{equation} 
\widetilde{\mathcal{C}}_l \ = \frac{l(l+1)}{8\pi^2} \int
d\Omega_{\q_1} \int d\Omega_{\q_2} \delT(\q_1) \delT(\q_2)
P_l(\q_1\cdot\q_2).
\end{equation}
Using the property of Gaussian random fields that,
\begin{eqnarray}
\langle \delT(\q_1) \delT(\q_2) \delT(\qp_1) \delT(\qp_2) \rangle &=&
\langle \delT(\q_1) \delT(\q_2) \rangle \langle \delT(\qp_1)
\delT(\qp_2) \rangle + \label{eq:gfield} \\ &&\langle \delT(\q_1)
\delT(\qp_1) \rangle \langle \delT(\q_2) \delT(\qp_2) \rangle +
\langle \delT(\q_1) \delT(\qp_2) \rangle \langle \delT(\qp_1)
\delT(\q_2) \rangle \nonumber
\end{eqnarray}
and eq.~(\ref{eq:corr}), we recover the well known result for full sky
CMB maps

\begin{equation} 
\ccov \ = \ \frac{2}{2l+1} \, \langle\mathcal{C}_l\rangle^2 \,
\delta_{l\lp} \ = \ \frac{2}{2l+1} \, \mathcal{C}_l^2 \, \delta_{l\lp}
\,,
\end{equation}
corresponding to $C_l$ being a sum of the squares of $2l+1$ Gaussian
variates, i.e. $\chi^2_{2l+1}$ distribution.  The measured power
spectrum at each multipole is independent (for full sky CMB maps).
The variance of the power spectrum estimator is not zero even in the
ideal case. Consequently, the measurement angular power spectrum from
the one available CMB sky map is inherently limited by an inevitable error
the \emph{Cosmic Variance}~\footnote{This is a direct
consequence of the sphere being compact and, consequently, an
inevitable, rigid lower bound on the uncertainty in the measurement of
angular power spectrum at a given multipole $l$. Otherwise, the effect
is the similar to the well-known sample variance of a finite
data-stream}.

\subsection{Circular Beam}

For measurements made with a circular beam, the temperature is a
linear transform of the actual temperature (see eq.~(\ref{eq:beam})).
So, it also represents a Gaussian random field. Hence,
eq.~(\ref{eq:gfield}) remains valid even for observed temperature
fluctuations. Moreover, the window function takes a simple form given
in eq.~(\ref{isowine}). Consequently, eq.~(\ref{eq:corr}) gets
modified to
\begin{equation} 
\langle \delTobs(\q_1) \delTobs(\q_2) \rangle \ = \ \sum_{l=0}^\infty
\frac{2l+1}{2l(l+1)} \, B_{l}^2 \, \mathcal{C}_l \, P_l(\q_1 \cdot
\q_2).
\end{equation}
The covariance matrix
\begin{equation}
\ccov \ = \ \frac{2}{2l+1} \, \cclobs^2 \delta_{l\lp}\ = \
\frac{2}{2l+1} \, (B_{l}^2 \mathcal{C}_l)^2 \, \delta_{l\lp},
\label{eq:Coviso}
\end{equation}
remains diagonal for circular beams, i.e., the measured power
spectrum at each multipole is independent of the power measured in the
other multipoles. The second equality follows from
eq.~(\ref{eq:Cliso}).

Including the instrumental noise spectrum in the measured power
spectrum $\mathcal{C}_l^N$, we obtain
\begin{equation} 
\ccovp \ = \ \frac{2\delta_{l\lp}}{2l+1} \left( \cclobs +
\mathcal{C}_l^N \right)^2,
\end{equation}
where we assume that the noise spectrum $\mathcal{C}_l^N$ is known
much better and, in particular, does not suffer from cosmic variance.
For the unbiased estimator given by eq.~(\ref{clcirub}), the well
known covariance matrix
\begin{equation} 
\mbox{Cov}(\club,\clubp) \ = \ B_l^{-4} \ccovp \ = \
\frac{2\delta_{l\lp}}{2l+1} \left( \mathcal{C}_l +
B_l^{-2}\mathcal{C}_l^N \right)^2
\end{equation}
is readily obtained from the linear transformation between
$\mathcal{C}_l^\prime$ and $\mathcal{C}_l^{UB}$~\cite{knox95,bonLH}.

%%%%%%%%%%%%%%%%%%%%%%%%%

\subsection{Non-circular Beam}

As expected, the covariance for the non-circular beam is considerably
more complicated.  We start with the general form of the two point
correlation function. Using eq.~(\ref{eq:clobs}), the general form of
the covariance matrix is
\begin{eqnarray}
\ccov &=& \frac{l\lp(l+1)(\lp+1)}{(4\pi)^4} \sum_{l_1,l_2=0}^{\infty}
\frac{(2l_1+1)(2l_2+1)}{l_1l_2(l_1+1)(l_2+1)} \, \mathcal{C}_{l_1}
\mathcal{C}_{l_2} \,\int d^4\Omega w(\q_1)w(\q_2)w(\qp_1)w(\qp_2)
\times \nonumber \\ && P_l(\q_1 \cdot \q_2) P_\lp(\qp_1 \cdot \qp_2)
\, [ W_{l_1}(\q_1,\qp_1)W_{l_2}(\q_2,\qp_2) \, + \,
W_{l_1}(\q_1,\qp_2)W_{l_2}(\q_2,\qp_1)],
\end{eqnarray}
where for brevity we denote $d^4\Omega \ \equiv \
d\Omega_{\q_1}d\Omega_{\q_2} d\Omega_{\qp_1}d\Omega_{\qp_2}$.  

Noting the interchangeability of the dummy variables $\qp_1$ and
$\qp_2$, we combine the two terms in the above equation to obtain
\begin{eqnarray}
\ccov &=& 2 \left[ \frac{l\lp(l+1)(\lp+1)}{(4\pi)^4} \right]
\sum_{l_1,l_2=0}^{\infty} \frac{(2l_1+1)(2l_2+1)}{l_1 l_2
(l_1+1)(l_2+1)} \, \mathcal{C}_{l_1} \mathcal{C}_{l_2} \, \times
\nonumber\\ &&\int d^4\Omega w(\q_1)w(\q_2)w(\qp_1)w(\qp_2) P_l(\q_1
\cdot \q_2) P_\lp(\qp_1 \cdot \qp_2)
W_{l_1}(\q_1,\qp_1)W_{l_2}(\q_2,\qp_2).
\end{eqnarray}
We expand the Legendre Polynomials in terms of spherical harmonics
(eq.~(\ref{eq:pl})) and use the expression for the window function in
eq.~(\ref{eq:WD}) to obtain
\begin{eqnarray}
&&\ccov = \frac{l\lp(l+1)(\lp+1)}{8\pi^2(2l+1)(2\lp+1)}
\sum_{l_1,l_2=0}^{\infty} \frac{(2l_1+1)(2l_2+1)}{l_1 l_2
(l_1+1)(l_2+1)} \, \mathcal{C}_{l_1} \mathcal{C}_{l_2} B_{l_1}^2
B_{l_2}^2 \sum_{m=-l}^{l} \sum_{m^\p=-\lp}^{\lp} \sum_{m_1=-l_1}^{l_1}
\sum_{m_2=-l_2}^{l_2} \nonumber\\ &&\left[
\sum_{m^\p_1,\mpp_1=-l_1}^{l_1} \B_{l_1m^\p_1} \B_{l_1\mpp_1}^{*} \int
d\Omega_{\q_1} w(\q_1) Y_{lm}^*(\q_1)
D_{m_1m^\p_1}^{l_1}(\q_1,\rho(\q_1)) \int d\Omega_{\qp_1}w(\qp_1)
Y_{\lp m^\p}(\qp_1)D_{m_1\mpp_1}^{l_1*}(\qp_1,\rho(\qp_1))
\right. \nonumber\\ && \left. \sum_{m^\p_2,\mpp_2=-l_2}^{l_2} \B_{l_2
m^\p_2}^* \B_{l_2\mpp_2}\int d\Omega_{\q_2}w(\q_2) Y_{lm}(\q_2) D_{m_2
m^\p_2}^{l_2*}(\q_2,\rho(\q_2)) \int d\Omega_{\qp_2} w(\qp_2) Y_{\lp
m^\p}^*(\qp_2)D_{m_2\mpp_2}^{l_2}(\qp_2,\rho(\qp_2)) \right]\,
,\nonumber\\
\label{ccovgen}
\end{eqnarray}
as the general expression for error covariance for angular power
spectrum for non-circular beams.  Note that even for full, uniform sky
observations, $w(\q)=1$, the error covariance matrix is no longer
diagonal.

To make further progress analytically, we restrict to the case of
uniform, full sky coverage ($w(\q)=1$) with no beam rotation
($\rho(\q)=0$). Using the integration of eq.~(\ref{eq:intYD}) and
after a considerable algebra we may write the expression for
covariance as
\begin{eqnarray}
\ccov = \frac{l\lp(l+1)(\lp+1)}{8} \sum_{m=-L}^{L} \left[
\sum_{l_1=|m|}^{\infty} B_{l_1}^2 \mathcal{C}_{l_1}
\frac{(2l_1+1)}{l_1(l_1+1)} \sum_{m^\p_1=-l_1}^{l_1} \B_{l_1m^\p_1}
I_{mm^\p_1}^{l l_1} \sum_{\mpp_1=-l_1}^{l_1} \B_{l_1\mpp_1}^{*}
I_{m\mpp_1}^{\lp l_1} \right]^2 \label{eq:covaniso}
\end{eqnarray}
where $L={\rm min}\{l,\lp\}$ is the smaller between $l$ and $\lp$. The
integrals $I_{mm^\p}^{l\lp}$ are defined in \S\ref{secBias} and the
analytical expressions for $m^\prime=0,\pm 2$ are given.  It is
straightforward to verify that the above equation correctly reproduces
the expression for the error-covariance in the circular beam case
given by eq.~(\ref{eq:Coviso}).

\begin{figure}
\centering
\includegraphics[width=0.75\textwidth,angle=0]{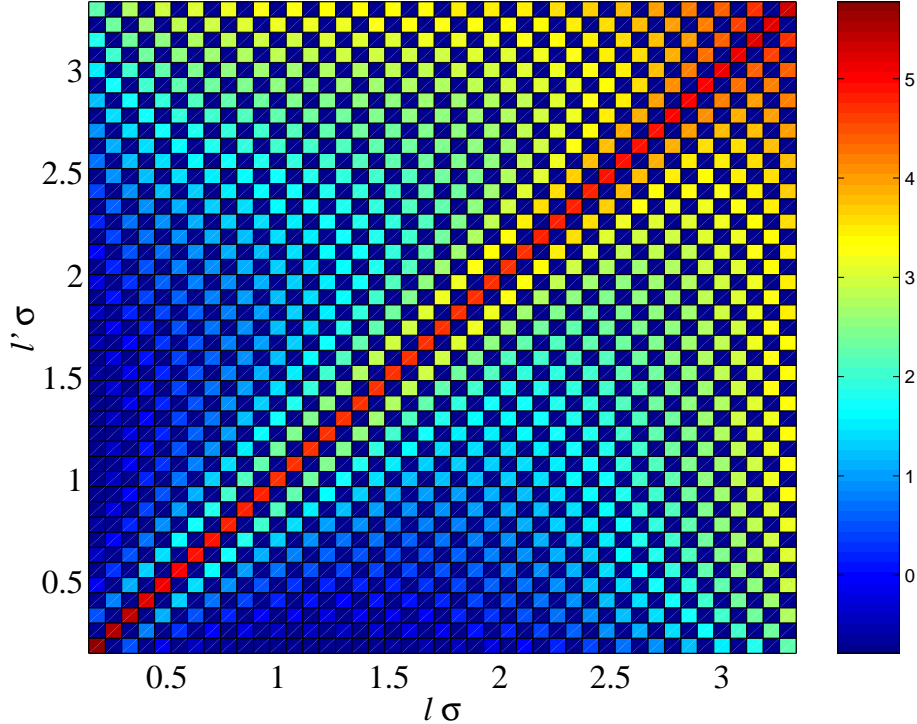}
\caption{Log of the normalized covariance matrix Cov$(C_l,
 C_{l'})/(B_lB_\lp)^2$ [in the units of $(\mu K)^4$] is plotted for an
 elliptical beam of eccentricity $e=0.6$ and mean beam-width
 $\bar\sigma = 0.074$. Due to the non-circularity of the beam, the
 error in CMB angular power spectrum estimate at different multipoles
 are no longer independent. We notice that the off-diagonal elements
 of the error covariance matrix are pronounced for $l \bar\sigma \geq 1$.}
   \label{fig:covmat}
\end{figure}

For evaluation of the covariance matrix, we note that though the
summation over $l_1$ runs from $0$ to $\infty$, the contributions are
significant only around $l \sim 1/\bar\sigma$ and the summation can be
truncated suitably.  Further, for most beams we can confine to the
leading order approximation as in eq.~(\ref{wigwin40}), by neglecting
all the $\B_{lm}$'s for $m\ge 4$. For mild deviations from circular
beams, the observed power spectrum at different multipoles are weakly
correlated ( $\sim \B_{l2}\B_{\lp 2}$). The error-covariance matrix
can be diagonalized to find the independent linear combinations
of estimators (eigenvectors), and the variances of theses independent
estimators are given by the corresponding eigenvalues. These
eigenvalues are necessarily larger that the cosmic variance
corresponding to a circular beam.

The inclusion of instrumental noise is similar to what was done in the
circular beam case.  The covariance
\begin{equation} 
\ccovp \ = \ \ccov \ + \ \frac{2\delta_{l\lp}}{2l+1} \left[
2\cclobs\mathcal{C}_l^N \, + \, (\mathcal{C}_l^N)^2 \right]
\label{eq:covp}
\end{equation}
clearly reproduces the result in eq.~(\ref{eq:Coviso}) in the limit of
a circular beam. Figure~\ref{fig:covmat} shows a density plot of the
elements of the covariance matrix for a non-circular (elliptical) beam
with no rotation. In contrast to the case for incomplete (cut) sky
case, where the effects are at small $l$ (see \cite{efs04}), the
non-circular beam affects the large multipoles region ($l\bar\sigma
\ge 1$). The pseudo-$C_l$ approach is close to optimal for large $l$
hence it may be more important to account for non-circular beams
effects than the cut-sky, since it is possible to use maximum
likelihood estimator for small $l$.

The error-covariance matrix for the unbiased estimator
eq.~(\ref{clncub}) for non-circular beams is given by
\begin{eqnarray}
\mbox{Cov}(\club,\clubp) &=& \sum_{l_1} \sum_{l_2} A_{ll_1}^{-1} A_{\lp l_2}^{-1}  \mbox{Cov}(\widetilde{\mathcal{C}}_{l_1}^\p,\widetilde{\mathcal{C}}_{l_2}^\p) \nonumber\\ &=& \sum_{l_1} \sum_{l_2} \alpha_{ll_1} \alpha_{\lp l_2} (B_{l_1}B_{l_2})^{-2} \left[ \ccov \ + \ \frac{2\delta_{l\lp}}{2l+1} \left\{ 2\cclobs\mathcal{C}_l^N \, + \, (\mathcal{C}_l^N)^2 \right\} \right],
\end{eqnarray}
where the matrix $\alpha_{l\lp} \equiv B_{\lp}^{-2} A_{l\lp}^{-1}$,
being very close to identity, demonstrates that the beam-modified
cosmic variance part of the covariance of unbiased estimator weakly
depends on $B_l$'s, whereas the noise part depends on them
significantly.

\section{Discussion and Conclusion}
\label{secConcl}

We present an analytic framework for addressing the effect of
non-circular experimental beam function in the estimation of the
angular power spectrum $\mathcal{C}_l$ of CMB anisotropy. Non-circular
beam effects can be modeled into the covariance functions in
approaches related to maximum likelihood estimation~\cite{max97,bjk98}
and can also be included in the Harmonic ring~\cite{harm02} and
ring-torus estimators~\cite{wan_han03}.  The latter is promising since
it reduces the computational costs from $N^3$ to $N^2$. However, all
these methods are computationally prohibitive for high resolution maps
and, at present, the computationally economical approach of using a
pseudo-$C_l$ estimator appears to be a viable option for extracting
the power spectrum at high multipoles~\cite{efs04}. The pseudo-$C_l$
estimates have to be corrected for the systematic biases.  While
considerable attention has been devoted to the effects of
incomplete/non-uniform sky coverage, no comprehensive or systematic
approach is available for non-circular beam.  The high sensitivity,
`full' (large) sky observation from space (long duration balloon)
missions have alleviated the effect of incomplete sky coverage and
other systematic effects such as the one we consider here have gained
more significance. Non-uniform coverage, in particular, the galactic
masks affect only CMB power estimation at the low multipoles. Recently
proposed hybrid scheme promotes a strategy where the power spectrum at
low multipoles is estimated using optimal Maximum Likelihood methods
and pseudo-$C_l$ are used for large multipoles.

We have shown that non-circular beam is an effect that dominates at
large $l$ comparable to the inverse beam width. For high resolution
experiment, the optimal maximum likelihood methods which can account
for non-circular beam functions are computationally prohibitive.  In
implementing pseudo-$C_l$ estimation, the non-circular beam effect
could dominate over the effects of more well studied effect of
non-uniform sky coverage. Our work provides a convenient approach for
estimating the magnitude of this effect in terms of the leading order
deviations from a circular beam. The perturbation approach is very
efficient. For most CMB experiments the leading few orders capture
most of the effect of beam non-circularity.  The perturbation approach has
allowed the development of computationally rapid method of computing
window functions~\cite{TR:2001}. Our work may similarly yield
computationally rapid methods correcting for beam non-circularity.

The quantitative estimates of the off-diagonal matrix elements of the
bias and error-covariance for `non-rotating' beam graphically
illustrate the general features that can be gleaned from our analytic
results. They show that the beam non-circularity affects the $C_l$
estimation on multipoles larger than the inverse beam width. A strong
dependence on the eccentricity of the beam is also seen. {\em We
caution against interpreting these results as a measure of the
non-circular beam effects for any real CMB experiment.} The analytical
results are limited to non-rotating beams and uniform sky coverage.
Numerical results do not include scan pattern of any known
experiment. Numerical calculations of the bias matrix for a `toy'
scanning strategy where the beam rotates on the sky indicates the
possibility of significant corrections. The bias due to non-uniform
sky coverage can have interesting coupling to the bias from beam
non-circularity. On the other hand, it has also been shown that
effects of non-circular beams can be diluted if the scan pattern is
such that each point in the sky is revisited by the beam with a
different orientation at different time~\cite{whu01}.  The numerical
implementation of our method can readily accommodate the case when
pixels are revisited by the beam with different
orientations. Evaluating the realistic bias and error-covariance for a
specific CMB experiment with non-circular beams would require
numerical evaluation of the general expressions for $A_{l\lp}$ in
eqs.~(\ref{eq:a}) using real scan strategy and account for
inhomogeneous noise and sky coverage. We defer such an exercise to
future work.

It is worthwhile to note in passing that that the angular power $C_l$
contains all the information of Gaussian CMB anisotropy only under the
assumption of statistical isotropy.  Gaussian CMB anisotropy map
measured with a non-circular beam corresponds to an underlying
correlation function that violates statistical isotropy. In this case,
the extra information present may be measurable using, for example,
the bipolar power spectrum~\cite{haj_sour03}. Even when the beam is
circular the scanning pattern itself is expected to cause a breakdown
of statistical isotropy of the measured CMB anisotropy
~\cite{master}. For a non-circular beam, this effect could be much more
pronounced and, perhaps, presents an interesting avenue of future study.

In addition to temperature fluctuations, the CMB photons coming from
different directions have a random, linear polarization. The
polarization of CMB can be decomposed into $E$ part with even parity
and $B$ part with odd parity.  Besides the angular spectrum
$C_l^{TT}$, the CMB polarization provides three additional spectra,
$C_l^{TE}$, $C_l^{EE}$ and $C_l^{BB}$ which are invariant under parity
transformations. The level of polarization of the CMB being about a
tenth of the temperature fluctuation, it is only very recently that
the angular power spectrum of CMB polarization field has been
detected. The Degree Angular Scale Interferometer (DASI) has measured
the CMB polarization spectrum over limited band of angular scales in
late 2002~\cite{kov_dasi02}. The WMAP mission has also detected CMB
polarization~\cite{kog_wmap03}. WMAP is expected to release the CMB
polarization maps very soon. Correcting for the systematic effects of
a non-circular beam for the polarization spectra is expected to become
important soon. Our work is based on the perturbation approach of
~\cite{TR:2001} which has been already been extended to the case of
CMB polarization~\cite{fos02}. Extending this work to the case CMB
polarization is another line of activity we plan to undertake in the
near future.

In summary, we have presented a perturbation framework to compute the
effect of non-circular beam function on the estimation of power
spectrum of CMB anisotropy. We not only present the most general
expression including non-uniform sky coverage as well as a non-circular
beam that can be numerically evaluated but also provide elegant
analytic results in interesting limits. In this work, we have skipped
over the effect of non-circular beam functions on map-making step.  In
simple scanning strategies, our results may be readily applied in this
context.  As CMB experiments strive to measure the angular power
spectrum with increasing accuracy and resolution, the work provides a
stepping stone to address a rather complicated systematic effect of
non-circular beam functions.

%%%%%%%%%%%%%%%%%%%%%%%%%%%%%%%%%%%%%%%%%%%%%%%%%%%%%%%%%%%%%%%%%%%%%%%%%%%

\appendix

\section{Elliptical Gaussian fit to the WMAP beam maps}
\label{wmapbeamfit}

We briefly describe an exercise in characterizing non-circular beams
in CMB experiments using the beam maps of the WMAP mission.  We
analyzed the WMAP raw beam images in the Q1, V1 and W1
\cite{LamdaBl,pag03} bands using two different standard software
packages. We use the elliptical Gaussian fit allowed by the well known
radio-astronomy software, AIPS and a more elaborate ellipse fitting
routine available within the standard astronomical image/data
processing software IRAF.  The ELLIPSE task in the STSDAS package of IRAF,
which uses the widely known ellipse fitting routines by
Jedrzejewski~\cite{iraf}, allows independent elliptical fits to the
isophotes. This significant greater degree of freedom in fitting to the
non-circular beam allows us to assess whether a simple elliptical
Gaussian fit is sufficient. The three bands see Jupiter in the two
horns (labeled A and B) as a point source.  The fitting routine fits
ellipses along iso-intensity contours of the beam image, parameterized
by position angle (PA), ellipticity ($\bar\epsilon$) and position of
the center. Each of these parameters can be independently varied.  The
distance between successive ellipses can also be independently
varied. The eccentricity $e$ is related to ellipticity $\bar\epsilon$
as $e = \sqrt{1 - (1-\bar\epsilon)^2}$ (Please see Table~\ref{tab:2}).

\begin{figure}[h]
\centering
\includegraphics[width=0.75\textwidth]{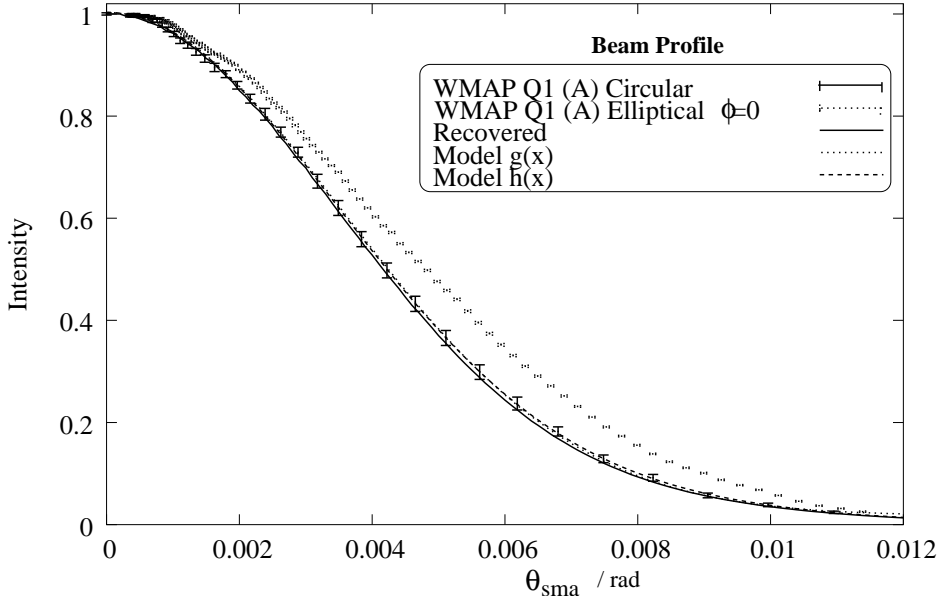}
\caption{The beam profile is characterized by intensity along the semi
major axis (SMA). The beam in Q-band for WMAP experiment was analyzed
using IRAF and fitted to both circular and elliptical profiles. We
have plotted the best fit circular profile (solid error bars) and
overlaid the profile recovered by inverting the WMAP beam transforms,
available at LAMBDA website (solid line). Two analytical models for
circular beam profile $g(\theta)$ and $h(\theta)$ are also considered,
and the best fit profiles are overlaid. We find that these models are
consistent with the IRAF and WMAP data. We have also plotted the best
fit elliptical profile along SMA (broken error bar). Notice that the
error bars in this case are much smaller than those for circular
profile, implying a better agreement with the data.}
\label{fig:btheta}
\end{figure}

We fit the the beams in two different ways: (a) by holding the
ellipticity constant to $\bar \epsilon = 0.05$ and freely varying the
position angle and center and (b) fixing the center to be the pixel
with the highest intensity (normalized to 1.0 at the central pixel)
and varying ellipticity and position angle. In the first case, we
get the closest approximation to circular beam profiles as used in
WMAP data analysis.  This beam has no azimuthal ($\phi$) dependence.
In the latter case, we get the elliptical profile of the WMAP beam
which depends on both the polar ($\theta$) and azimuthal ($\phi$)
distance from the pointing direction. Notice that in this case it is
sufficient to provide the intensity along a particular direction
(usually, the semi-major axis or $\phi=0$) and the ellipticity $\bar\epsilon$.

\begin{figure}[h]
\includegraphics[width=0.6\textwidth,angle=-90]{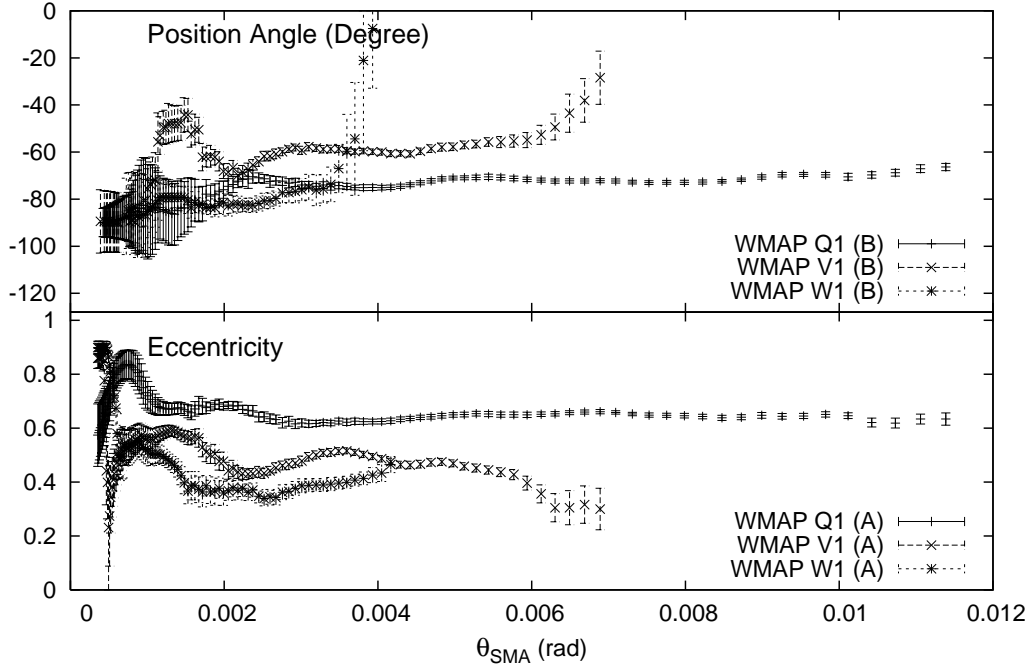}
\caption{The beams for WMAP experiment in three bands Q, V and W for
both the horns (A and B) were fitted to elliptical profiles using
IRAF. The plot above shows the fitted eccentricity and position angle
along the semi major axis (SMA).  The presence of side-bands in the
smaller beams (W band) makes it difficult for IRAF to model them
sufficiently well.  However, in the Q1 band, such sub-structures in
the beam are not present thus allowing the IRAF ellipse fitting
routine to fit reasonably good ellipses which have consistent
eccentricities ($e \sim 0.65$) and position angles all along the
SMA. The V1 beam is smaller in extent than the Q1 beam and its
eccentricity was determined to be $e \sim 0.46$ using IRAF. The
highest resolution beam in in W1 band, whose eccentricity was
determined to be $e \sim 0.40$.}
\label{fig:eccANDpa}
\end{figure}

Even a visual inspection reveals that the Q1 beam map plotted in
Fig~\ref{fig:beam} is non-circular and the iso-intensity
contours distinctly elliptical. Thus it comes as no surprise that the
error bars as shown in Figure~\ref{fig:btheta} for circularized beam
are larger than those for the elliptical profile.  As a consistency
check, we take the WMAP Q1 beam transfer function $B_l$ from WMAP
first year data archived at publicly available LAMBDA
site~\cite{LamdaBl} and `recover' the circular beam profile
$B(\theta)$ using eq.~(\ref{eq:BthetafromBl}).

From Figure~\ref{fig:btheta}, it is clear that this `recovered' beam
profile is in good agreement with that obtained by IRAF.  This allows
us to make some statements about the profile fitting in CMB
experiments, in the context of WMAP beams. The beam profile
$B(\theta)$ has been modeled as a Gaussian times a sum of even order
Hermite polynomials ($H_{2n}$) by the WMAP team~\cite{pag03}. To
compare, we have also modeled the beam profile with a function
$h(\theta)$ given by
\begin{equation}
h(\theta) = \exp \left (-\frac{1}{2} \ \alpha \theta^2 \right ) \left
( h_0 + h_2 H_2(\theta) + h_4 H_4(\theta) \right ),
\end{equation} 
where $\alpha, h_0, h_2$ and $h_4$ are unknown parameters to be fixed
by least squared method. We found that this model fits the data very
well with a reduced $\chi^2$ of about $0.7$. However, on closer
analysis, it is found that the chief role of the Hermite polynomials
is to add a constant baseline over and above the Gaussian. To test
this hypotheses, we choose another form of the fitting function
$g(\theta)$ given by
\begin{equation} 
g(\theta) = g_0 + g_1 \exp \left (-\frac{1}{2}\ g_2 \theta^2 \right ), 
\end{equation}
where $g_0, g_1$ and $g_2$ are parameters of the model. It is very
interesting to note that this model also fits the data very well with
a reduced $\chi^2$ of about $0.8$ for the best fitted parameters. In
all fairness, $g(\theta)$ serves as a simpler model for the beam
profile. We cannot point to the precise origin for the
baseline. However, such `skirts' in beam responses
are not uncommon in radio-astronomy. At this point, our observation
should perhaps merit a curious aside, if not as an alternative
approach to beam modeling.  Our best-fit models $g(\theta)$ and
$h(\theta)$, along with the IRAF fitted data points to the WMAP Q1 (A)
beam is shown in Figure~\ref{fig:btheta}.

As shown earlier in this paper, the effects of non-circularity of the
CMB experimental beams show up in the power-spectral density estimates
through the off-diagonal elements of the bias matrix $A_{ll'}$. As
shown in eq.~(\ref{eq:a}), these in turn can be expressed in terms of
the leading components of the harmonic transform of the beam. In
general the harmonic decomposition of a non-circular beam may
have to be done numerically. But for the
particular case of an elliptical Gaussian beam, a closed form
expression given by eq.~(\ref{eq:blm:elliptical})
serves as a useful test-bed for us.  Thus another motivation for
fitting ellipses to WMAP beams using IRAF was to get a handle on the
eccentricity of these beams so as to find the harmonic transform
components of an elliptical Gaussian beam of similar
eccentricity. This allows us to give more realistic estimates of the
effect of non-circularity of the beam on $C_l$ estimates.

\begin{table}[h]
\caption{The result of ellipse fitting using IRAF on the Q1, V1 and W1
beams of the WMAP experiments. The frequency quoted is the `effective'
frequency of the corresponding band from Page et. al.~\cite{pag03}. The
presence of sub-structures in the W1 band makes it difficult to fit
elliptical contours to the beam.}

\begin{center}
\begin{tabular}{lllc}
\hline
Beam               & Frequency     & Eccentricity        & Position Angle              \\
                   & (GHz)         &                     & (degree)                    \\ \hline \hline
Q1 (A)             & 40.9          & 0.65                & +80                         \\
Q1 (B)             & 40.9          & 0.67                & -80                         \\
V1 (A)             & 60.3          & 0.48                & +60                         \\
V1 (B)             & 60.3          & 0.45                & -60                         \\
W1 (A)             & 93.5          & 0.40                & ---\footnotemark                           \\ \hline
\label{tab:1}
\footnotetext{Not well determined by the IRAF ellipse fitting
routine.}
\end{tabular}
\end{center}
\end{table}

It is interesting to note how the fitted eccentricities vary as a
function of the distance along the semi-major axis of the fitted
ellipses for various beams.  The smaller beams (V1 and W1) have
sufficient sub-structure in the form of side-lobes which throws the
ellipse fitting routine off course. However, where the sub-structure
is less pronounced, we find that the eccentricities of the fitted
ellipses takes a constant value. Toward the center of the ellipses,
there are far too few pixels to average over, which in turn manifests
as large error bars in the eccentricities and position angles of the
ellipses. In Figure~\ref{fig:eccANDpa}, we notice that the Q1 beam has
a very elliptical profile with eccentricity $e \gsim 0.65$ and
position angle of about $75^\circ$. We also fitted the Q1 (A) beam to
an elliptical Gaussian model using radio astronomy standard data
analysis software AIPS and got consistent numbers for the
eccentricity. However the IRAF modeling gives us more freedom to vary
the eccentricity and position angle as we move away from the center of
the ellipse and the result is that the beam is modeled more
accurately.

%%%%%%%%%%%%%%%%%%%%%%%%%%%%%%%%%%%%%%%%%%%%%%%%%%%%%%%%%%%%%%%%%%%%%%%%%%%

\section{Details of analytic derivations}
\label{Appint}
\noindent

In the appendix we provide the details of the analytical steps
involved in deriving some of the expressions used in the main text.
This is designed to keep the paper self contained and easy to extend.

First, we outline the steps involved in evaluating the integral
$I_{02}^{l\lp}+I_{0-2}^{l\lp} = 2\int_{-1}^{1} \,
d_{00}^l(\theta) d_{02}^{l^\prime}(\theta) d\cos\theta$ to obtain the
result in eq.~(\ref{eq:Im0}).

Using the expressions~\cite{TR:2001,VMK} for $d_{00}^l$ and $d_{02}^l$
in terms of Legendre Polynomials and its derivatives,
\begin{equation} 
\int_{-1}^{1} \, d_{00}^l(\theta) d_{02}^{l^\prime}(\theta)
d\cos\theta \ = \ -\frac{\lp(\lp+1)}{\kappa}\int_{-1}^{1} \, P_l(x)
P_{\lp}(x) dx \, + \, \frac{2}{\kappa} \int_{-1}^{1} \, x P_l(x)
P_{\lp}^\prime(x) dx,
\end{equation}
where, $\kappa \ \equiv \ \sqrt{(l-1)l(l+1)(l+2)}$.  The first
integral is simply the orthogonality of Legendre polynomials
\begin{equation}
\int_{-1}^{1} \, P_l(x) P_{\lp}(x) dx \ = \
\frac{2\delta_{l\lp}}{2l+1}. \label{eq:plortho}
\end{equation}
Further, we can show that for odd values of $l+l^\prime $,
\begin{equation}
\int_{-1}^{1} \, x \, P_l(x) \, P_{l^\prime}^\prime(x) \, dx \ = 0,
\end{equation}
and for  even values of $l+l^\prime$,
\begin{equation}
\int_{-1}^{1} \, x \, P_l(x) \, P_{l^\prime}^\prime(x) \, dx \ =
\left\{ \begin{array}{cl} 2 & \mbox{if $l<l^\prime$}\\ 0 & \mbox{if
$l>l^\prime$}\\ 2l/(2l+1) & \mbox{if $l=l^\prime$}. \end{array}
\right.
\end{equation}
Assembling all these we can derive  eq.~(\ref{eq:Im0}).

Next we evaluate the more general integral
$I_{m2}^{l\lp}+I_{m-2}^{l\lp} \equiv \int_{-1}^{1} d_{m0}^l(\theta) \,
[d_{m2}^{\lp}(\theta) + d_{m-2}^{\lp}(\theta)] \, d\cos\theta$ to
obtain the expression in eq.~(\ref{eq:Im2}). The first step is to
express $d_{m\pm2}^l(\theta)$ in terms of $d_{m0}^{l^\prime}(\theta)$.
Using the recurrence relations for Wigner~$D$ functions (see eq.~(4)in
\S4.8.1,\cite{VMK}) and using the fact that
\begin{equation}
D_{mm^\p}^l(\phi,\theta,\rho) \ = \ e^{-im\phi} \, d_{mm^\p}^l(\theta)
\, e^{-im^\p\rho}
\end{equation}
we get the recurrence relations for Wigner-$d$ functions:
\begin{eqnarray}
\sin\theta \, d_{mm^\p+1}^l(\theta) &&= \
\frac{\sqrt{(l^2-m^2)(l+m^\p)(l+m^\p+1)}}{l(2l+1)}d_{mm^\p}^{l-1}(\theta)\\
&-& \frac{m\sqrt{(l-m^\p)(l+m^\p+1)}}{l(l+1)}d_{mm^\p}^l(\theta) \ - \
\frac{\sqrt{[(l+1)^2-m^2](l-m^\p)(l-m^\p+1)}}{(l+1)(2l+1)}d_{mm^\p}^{l+1}(\theta). \nonumber
\end{eqnarray}
Using these relations for $d_{m2}^l$ we may write,
\begin{equation}
d_{m2}^l(\theta) \ = \ \frac{\kappa}{\sin^2(\theta)} \,
[\kappa_0d_{m0}^l(\theta) + \kappa_1d_{m0}^{l+1}(\theta) +
\kappa_{-1}d_{m0}^{l-1}(\theta) + \kappa_2d_{m0}^{l+2}(\theta) \, + \,
\kappa_{-2}d_{m0}^{l-2}(\theta)],
\end{equation}
where
\begin{eqnarray*} 
\kappa_0 \ &\equiv& \ \frac{m^2}{l^2(l+1)^2} \, - \,
\frac{l^2-m^2}{l^2(4l^2-1)} \, - \,
\frac{(l+1)^2-m^2}{(l+1)^2(2l+1)(2l+3)}, \\ \kappa_1 \ &\equiv& \
2m\frac{\sqrt{(l+1)^2-m^2}}{l(l+1)(l+2)(2l+1)}, \ \ \kappa_{-1} \
\equiv \ -2m\frac{\sqrt{l^2-m^2}}{l(l^2-1)(2l+1)},\\ \kappa_2 \
&\equiv& \
\frac{\sqrt{[(l+1)^2-m^2][(l+2)^2-m^2]}}{(l+1)(l+2)(2l+1)(2l+3)} \ \
\mbox{and} \ \ \kappa_{-2} \ \equiv \
\frac{\sqrt{(l^2-m^2)[(l-1)^2-m^2]}}{l(l-1)(4l^2-1)}.
\end{eqnarray*}
Also, since $d_{m-2}^l(\theta) = (-1)^{l+m}d_{m2}^l(\pi-\theta)$ and
$d_{m0}^l(\pi-\theta) = (-1)^{l+m}d_{m0}^l(\theta)$ we can write,
\begin{equation}
d_{m-2}^l(\theta) \ = \ \frac{\kappa}{\sin^2(\theta)} \, [\kappa_0d_{m0}^l(\theta) - \kappa_1d_{m0}^{l+1}(\theta) - \kappa_{-1}d_{m0}^{l-1}(\theta) + \kappa_2d_{m0}^{l+2}(\theta) \, + \, \kappa_{-2}d_{m0}^{l-2}(\theta)].
\end{equation}

Using the expression for $d_{m\pm 2}$ we can make the following
substitution
\begin{equation} 
d_{m2}^l(\theta)+d_{m-2}^l(\theta) \ = \ 2\kappa \,
[\kappa_0d_{m0}^l(\theta) + \kappa_2d_{m0}^{l+2}(\theta) \, + \,
\kappa_{-2}d_{m0}^{l-2}(\theta)]/\sin^2\theta\,,
\end{equation}
in the integral we seek to evaluate.  We use the following integral
for $l \le l^\p$ and $L={\rm min}\{l,\lp\}\ge |m|>0$,
\begin{equation}
\int_{-1}^{1} \, d_{m0}^l(\theta) d_{m0}^{l^\p}(\theta)
\frac{d\cos\theta}{\sin^2\theta} \ = \ \left\{ \begin{array}{ll}
\frac{1}{|m|}\sqrt{\frac{(l+|m|)!(l^\p-|m|)!}{(l-|m|)!(l^\p+|m|)!}} &
\mbox{even $l+l^\p $}\\ \\ 0 & \mbox{{odd $l+l^\p$}}\end{array}
\right. \label{eq:intdd}
\end{equation}
and obviously, for $l > l^\p$, $l$ and $l^\p$ have to be interchanged
in the above expression. We then obtain $I_{m2}^{l\lp}+I_{m-2}^{l\lp}$
as given in eq.~(\ref{eq:Im2}).

The integral in eq.~(\ref{eq:intdd}) can also be readily derived.  We
use the fact that
\begin{equation}
d_{m0}^l(\theta) \ = \ (-1)^m \, \sqrt{\frac{(l-m)!}{(l+m)!}} \,
P_l^m(\cos\theta)\,,
\end{equation}
which leads to
\begin{equation}
\int_{-1}^{1} \, d_{m0}^l(\theta) d_{m0}^{\lp}(\theta) \frac{d\cos\theta}{\sin^2\theta} \ = \ \sqrt{\frac{(l-m)!(\lp-m)!}{(l+m)!(\lp+m)!}} \, \int_{-1}^{1} P_l^m(x) P_{\lp}^m(x) \frac{dx}{1-x^2}\,. \label{eq:A1}
\end{equation}
The symmetry of Associated Legendre Polynomials, $P_l^m(-x) =
(-1)^{l+m}P_l^m(x)$ dictates that the integrand is antisymmetric for
odd values of $l + l^\prime$, hence the integral is zero.  However for
even values of $l + l^\prime $, we can evaluate the integral in the
following manner. One of the recurrence relations for Associated
Legendre Polynomials is (\cite{Arfken}, \S 12.5.)
\begin{equation}
P_l^m(x) \, = \, P_{l-2}^m(x) \, + \, (2l-1) \sqrt{1-x^2}
P_{l-1}^{m-1}(x)\,. \label{eq:A2}
\end{equation}
Using equation~(\ref{eq:A2}) we can write,
\begin{equation} 
\int_{-1}^{1} \, P_l^m(x) P_{l^\prime}^m(x) \frac{dx}{1-x^2} \ = \ \
\int_{-1}^{1} \, P_l^m(x) P_{l^\prime-2}^m(x) \frac{dx}{1-x^2} \, + \,
(2l^{\prime}-1)\int_{-1}^{1} \, P_l^m(x) P_{l^\prime-1}^{m-1}(x)
\frac{dx}{\sqrt{1-x^2}}\,. \label{eq:A3}
\end{equation}
We have provided a proof that the second integral on the right is zero
at the end of this section. Thus, from eq.~(\ref{eq:A3}) we have
\begin{equation}
\int_{-1}^{1} \, P_l^m(x) P_{l^\prime}^m(x) \frac{dx}{1-x^2} \ = \
\int_{-1}^{1} \, P_l^m(x) P_{l^\prime-2}^m(x) \frac{dx}{1-x^2}\,.
\end{equation}
In this way we can keep reducing $\lp$ by two each time until it
equals with $l$ (since $l+l^\prime$ is even and $l < \lp$ it
\emph{will} reduce to $l$). Thus, we have shown that,
\begin{equation}
\int_{-1}^{1} \, P_l^m(x) P_{l^\prime}^m(x) \frac{dx}{1-x^2} \ = \
\int_{-1}^{1} \, \left[ P_l^m(x) \right]^2 \frac{dx}{1-x^2} \ = \
\frac{1}{m} \frac{(l+m)!}{(l-m)!}\,, \label{eq:intPlm}
\end{equation}
where the second equality follows from the evaluation of a standard
integral, which can be obtained, for example, from
\cite{Arfken}. Substituting in eq.~({\ref{eq:A1}) we can evaluate the
integral for $m>0$. Clearly eq.~(\ref{eq:intPlm}) is valid for $l =
\lp$. For $l > \lp$, $l$ should be replaced by $\lp$ in that
equation. Moreover, using the property $d_{-m0}^l(\theta) = (-1)^m
d_{m0}^l(\theta)$, we can express the integral for any $m\ne 0$, as
given in eq.~(\ref{eq:intdd}).

Finally we prove the result used in simplifying eq.~(\ref{eq:A3}) that
for even values of $l+l^\prime$ and $l<l^\prime$,
\begin{equation} 
\int_{-1}^{1} \, P_l^m(x) P_{l^\prime-1}^{m-1}(x)
\frac{dx}{\sqrt{1-x^2}} \ = \ 0.
\end{equation}

Using the recurrence relation of Legendre Polynomials in
eq.~(\ref{eq:A2}), we can write
\begin{equation} 
\int_{-1}^{1} \, P_l^m(x) P_{l^\prime-1}^{m-1}(x)
\frac{dx}{\sqrt{1-x^2}} \ = \ \int_{-1}^{1} \, P_{l-2}^m(x)
P_{l^\prime-1}^{m-1}(x) \frac{dx}{\sqrt{1-x^2}} + (2l-1)\int_{-1}^{1}
\, P_{l-1}^{m-1}(x) P_{l^\prime-1}^{m-1}(x) dx\,. \label{eq:A13}
\end{equation}
Then from the orthogonality relation of associated Legendre Polynomials,
\begin{equation}
\int_{-1}^{1} \, P_{l}^{m}(x) P_{l^\prime}^m(x) dx \ = \
\frac{2}{2l+1} \, \frac{(l+m)!}{(l-m)!} \, \delta_{l l^\prime}\,\,,
\end{equation}
we can see that the second integral on the right of eq.~(\ref{eq:A13})
vanishes for $l^\prime \ne l$. Thus we have,
\begin{equation}
\int_{-1}^{1} \, P_l^m(x) P_{l^\prime-1}^{m-1}(x)
\frac{dx}{\sqrt{1-x^2}} \ = \ \int_{-1}^{1} \, P_{l-2}^m(x)
P_{l^\prime-1}^{m-1}(x) \frac{dx}{\sqrt{1-x^2}}\,.
\end{equation}
We can use the above equation iteratively since the lower indices of
the $P_l^m$'s will never match as $l^\prime>l$.  So the lower index of
the first polynomial in the integration can be reduced to either $m$
or $m+1$ (depending on $l-m$ is even or odd) by repeated use of the
above identity. Thus we may write
\begin{equation}
\int_{-1}^{1} \, P_l^m(x) P_{l^\prime-1}^{m-1}(x)
\frac{dx}{\sqrt{1-x^2}} \ = \ \left\{ \begin{array}{c} \int_{-1}^{1}
\, P_m^m(x) P_{l^\prime-1}^{m-1}(x) \frac{dx}{\sqrt{1-x^2}} \\ or\\
\int_{-1}^{1} \, P_{m+1}^m(x) P_{l^\prime-1}^{m-1}(x)
\frac{dx}{\sqrt{1-x^2}}\,.
\end{array} \label{eq:A16} \right.
\end{equation}
Finally using the relations,
\begin{eqnarray}
P_m^m &=& (-1)^m \, (2m-1)!! \, (1-x^2)^{m/2} \, = \,
(-1)(2m-1)\sqrt{1-x^2}P_{m-1}^{m-1}, \\ P_{m+1}^m &=& x(2m+1)P_m^m =
x(2m+1)(-1)(2m-1)\sqrt{1-x^2}P_{m-1}^{m-1} =
(-1)(2m+1)\sqrt{1-x^2}P_m^{m-1}\,,
\end{eqnarray}
and the orthonormality condition in eq.~(\ref{eq:plortho}) we can see
that in both the cases right side of eq.~(\ref{eq:A16}) is zero. This
completes the proof.

%%%%%%%%%%%%%%%%%%%%%%%%%%%%%%%%%%%%%%%%%%%%%%%%%%%%%%%%%%%%%%%%%%%%%%%%%%%

%%%%%%%%%%%%%%%%%%%%%%%%%%%%%%%%%%%%%%%%%%%%%%%%%%%%%%%%%%%%%%%%%%%%%%%%%%%
\acknowledgments 

S.M. and A.S.S. would like to acknowledge the financial assistance
provided by CSIR in the form of senior research fellowship. We
acknowledge the use of the High Performance Computing facility of
IUCAA. We would like to thank Jayaram Chengalur, Joydeep Bagchi and
Tirthankar Roychoudhury for very useful discussions at various stages
of this work. TS thanks Simon Prunet, Olivier Dore and Lyman Page for
encouraging discussions and useful comments. We are grateful to
Sudhanshu Barwe, C.~D.~Ravikumar and Amir Hajian for help. We would
like to thank Sanjeev Dhurandhar for his encouragement and support.

%%%%%%%%%%%%%%%%%%%%%%%%%%%%%%%%%%%%%%%%%%%%%%%%%%%%%%%%%%%%%%%%%%%%%%%%%%%

%%%%%%%%%%%%%%%%%%%%%%%%%%%%%%%%%%%%%%%%%%%%%%%%%%%%%%%%%%%%%%%%%%%%%%%%%%%

\end{document}